\documentclass[preprint,trackchanges]{aastex631}

\newcommand{\rom}[1]{\MakeUppercase{\romannumeral#1}}
\usepackage{comment}
\usepackage{hyperref}
\usepackage{amsmath}
\usepackage{cleveref}
\shorttitle{LAMOST-MRS Be stars}
\shortauthors{Wang et al.}

\begin{document}

\title{Identification of new classical Be stars from the LAMOST MRS survey}

\correspondingauthor{Luqian Wang, Jiao Li, You Wu}
\email{wangluqian@ynao.ac.cn, lijiao@bao.ac.cn, wuyou@nao.cas.cn}

\author[0000-0003-4511-6800]{Luqian Wang}
\affiliation{Yunnan Observatories, CAS, P.O. Box 110, Kunming 650011, Yunnan, China}

\author[0000-0002-2577-1990]{Jiao Li}
\affiliation{Key Laboratory of Space Astronomy and Technology, National Astronomical Observatories, \\
Chinese Academy of Sciences \\
Beijing 100101,China}

\author[0000-0002-3616-9268]{You Wu}
\affiliation{Key Laboratory of Space Astronomy and Technology, National Astronomical Observatories, \\
Chinese Academy of Sciences \\
Beijing 100101,China}

\author[0000-0001-8537-3583]{Douglas R.\ Gies}
\affiliation{Center for High Angular Resolution Astronomy and Department of Physics and Astronomy, \\
Georgia State University, P.O. Box 5060\\
 Atlanta, GA 30302-5060, USA}

\author[0000-0002-7420-6744]{Jin Zhong Liu}
\affiliation{Xinjiang Astronomical Observatory, Chinese Academy of Sciences, People's Republic of China}
\affiliation{School of Astronomy and Space Science, University of Chinese Academy of Sciences, 
Beijing 100049, People's Republic of China}

\author[0000-0002-1802-6917]{Chao Liu}
\affiliation{Key Laboratory of Space Astronomy and Technology, National Astronomical Observatories, \\
Chinese Academy of Sciences \\
Beijing 100101,China}

\author[0000-0001-9989-9834]{Yanjun Guo}
\affiliation{Yunnan Observatories, CAS, P.O. Box 110, Kunming 650011, Yunnan, China}
\affiliation{School of Astronomy and Space Science, University of Chinese Academy of Sciences, \\
Beijing 100049, China}

\author[0000-0001-5284-8001]{Xuefei Chen}
\affiliation{Yunnan Observatories, CAS, P.O. Box 110, Kunming 650011, Yunnan, China}
\affiliation{Center for Astronomical Mega-Science, Chinese Academy of Science, \\
Beijing 100012, China}

\author[0000-0001-9204-7778]{Zhanwen Han}
\affiliation{Yunnan Observatories, CAS, P.O. Box 110, Kunming 650011, Yunnan, China}
\affiliation{Center for Astronomical Mega-Science, Chinese Academy of Science, \\
Beijing 100012, China}



\begin{abstract}
Be stars are B-type main-sequence stars that display broad Balmer emission lines in their spectra. Identification of Be population is essential
to further exam the formation and evolutionary models. We report the 
detection of classical Be (CBe) stars from observations with the Large sky Area Multi-Object fiber Spectroscopic Telescope Medium Resolution Survey of Date Release 7 (LAMOST MRS DR7). We used a deep convolutional neural network, the {\tt ResNet} with an 18-layer module to
examine the morphology of the H$\alpha$ profile. We identified 1,162 candidate Be stars from the collection of 2,260,387 spectra 
for 789,918 stars in the database. The {\tt ResNet} network achieves a Be star classification accuracy of 99.5$\%$. Among the detections, 
151 of these are prior known Be stars cross-matched from the literature. By applying a three-step test, we identified 183 new 
CBe stars. We find that 41 CBe stars are members of known open clusters. Based upon an investigation of the 
kinematics of the identified CBe stars from the {\emph Gaia} EDR3 astrometric solutions, we identified 16 new runaways. These new 
identifications will provide as a reference for future follow-ups to further investigate their physical properties.        
\end{abstract}



\section{Introduction} \label{sec:intro}
Be stars are B-type main-sequence stars, and their optical spectra display or have displayed Balmer emission features that are formed in a 
decretion circumstellar disk around the star. Based on a study from optical interferometric and spectropolarimetric observations for a sample
of seven Be stars by \citet{Quirrenbach1997}, they suggest that the disks are geometric thin. The disk is now widely accepted as moving in a 
Keplerian motion and governed by viscosity \citep{rivinius2013}. \citet{Lee1991} proposed the viscous decretion disk model and suggested that 
materials are ejected from the central star to form a decretion circumstellar disk as a result of outward transferring angular momentum. Observational
works conducted in multi-wavelength regimes have successfully corroborated the disk properties as predicted by the model. Examples 
include a combination of photometric and spectroscopic study spans from near-UV to mid-IR to investigate the disk evolution of $\omega$ CMa 
\citep{Ghoreyshi2021}, the investigation of the disk density distribution and disk mass decretion rates for 80 Be stars from the infrared disk continuum 
emission \citep{Vieira2017}, and radio observations for 57 Be stars carried out by \cite{Klement2019} to monitor the spectral energy distribution of the disks. 

The portrait illustrates the formation mechanism for the Be phenomenon is far from complete, many formation scenarios have been 
suggested to fulfill the missing gaps, these include the rapid rotation, non-radial pulsation, and magnetic fields. 
Be stars are rapidly rotating stars, and their projected rotational velocities can reach up to $\sim80\%$ of their critical rotational velocities 
\citep{rivinius2013}. The origin of the rapid rotation of Be stars is still not fully understood, and growing evidence 
suggests that the rapid rotation is the result of mass and angular momentum accretion in a post, close binary interaction. The detections of Be stars 
with a compact companion are reported by several works, such as Be X-ray binary systems consisting of a neutron star companion 
\citep{Reig2011}, and a Be and stellar-mass black hole binary \citep{Casares2014}. At the low mass companion end, Be binary systems containing an 
evolved subdwarf companion have recently been detected through FUV spectroscopic investigations 
\citep{Peters2008,Peters2013,Peters2016,Wang2018,Wang2021}. Kinematics studies of Be stars from \citet{Berger2001} and \citet{Boubert2018} 
suggest that most of them are products of binary systems that have gone through past mass transfer. 

Many Be stars often show short-term variability and periodicity from spectroscopic and photometric observations. Non-radial pulsation
and presence of magnetic fields may also link to the Be phenomenon \citep{rivinius2013}. \citet{Semaan2011,Semaan2013} analyzed the light curves 
for Be stars in the CoRoT fields and reported that all Be stars are pulsators displaying multi-periodicities. The pulsation is also related to the 
circumstellar activities, such as disk build-ups, dissipation, and outburst \citep{Stefl2003,Richardson2021}. The presence of magnetic fields has also 
been suggested to account for the spectroscopic and photometric variabilities and modulations in Be stars. Theoretical
models, such as the magnetically torqued disk model by \citet{Brown2004} suggests that the magnetic fields may be accounted for the stellar wind 
mass loss and angular momentum transferring from the central star to the disk, and the magnetic rotator model by \citet{Maheswaran2003} argues that 
the magnetic fields may be responsible for the formation of the Be Keplerian disk. However, no large-scale magnetic fields have been observed yet, and 
weak magnetic fields were reported only in a few Be stars \citep{Hubrig2006,Hubrig2009}. 
Be stars are great laboratories for studying disk physics, such as disk formation and evolutionary processes. They are testbeds for investigating 
the star-disk interaction process. Identification for the Be population in the Galaxy is also helpful to constrain the models of formation channels for these
rapidly rotating early-type stars \citep{Shao2014}. 

Since the spectrographic observations of Be stars by \citet{struve1931}, many searches for Be stars have been carried out through
photometric or spectroscopic measurements of their emission features. \citet{jaschek1982} compiled a catalog of Galactic 1,159 Be 
stars from various literature. \citet{witham2008} presented a catalog of 4,853 H$\alpha$ emission point sources from the INT Photometric H$\alpha$ 
Survey using the Wide Field Camera (IN/TWFC) for the Northern Galactic Plane (IPHAS), and a later assessment conducted by \citet{Gkouvelis2016} 
suggests that about 70\% of the emission-line objects are classical Be stars (CBe). Utilizing the data from the Very Large Telescope Survey Telescope 
Photometric H$\alpha$ Survey (VPHAS+), \citet{Mohr-Smith2017} found a total of 14,900 massive OB stars in the Carina Arm region. 
\citet{aidelman2020} applied a neural network approach to classify 248 CBe stars from this database. Also, using different machine learning classifiers, 
\citet{Perez-Ortiz2017} identified 50 CBe candidate stars in the outskirt of the Large Magellanic Cloud from the Optical Gravitational Lensing 
Experiments (OGLE) {\emph I-} band light curves. Recently, \citet{vioque2020} applied a machine-learning approach to search for new Herbig Ae/Be stars 
using the photometric data from {\it Gaia} DR2. As a side product, they also reported the detection of 693 newly classified candidate CBe stars distributed over the Galactic plane. The Be Star Spectra (BeSS) database \citep{neiner2011} is a continuously updated spectral library contributed 
from both amateur and professional communities, and it contains optical spectra for about 2,000 CBe stars residing in the Milky Way and Magellanic 
Clouds (at the time of writing). Based upon observations made from the large surveys, such as the Apache Point Galactic Evolution Experiment 
(APOGEE), \citet{chojnowski2015} reported detections of 128 newly found CBe stars. Thanks to the large field-of-view of the Large sky Area Multi-Object 
fiber Spectroscopic Telescope (LAMOST), \citet{Lin2015} reported the detection of 192 Galactic CBe candidates stars based upon the low-
resolution spectra from the DR1, and \citet{hou2016} reported the discovery of 5,603 new CBe stars from the DR2.   

Motivated by the recent release of a large collection of spectroscopic observations from the LAMOST DR7, we are interested in identifying CBe 
stars from this large and homogeneous database. Here we report our work of using a deep convolutional neural network approach to classify CBe 
stars from the Date Release 7 of the medium-resolution spectra database. The structure of this work is as follows. We report the data collected from the 
LAMOST medium resolution survey in Section~\ref{sec:observations}. In Section~\ref{sec:elimination}, we describe the procedures to select the 
early type OB stars from the DR7 database. We report the details for applying the deep convolutional neural network method {\tt ResNet} to classify 
Be candidate stars and describe their morphological features in Section~\ref{sec:identification_Be}. The work of identification of CBe stars from the 
preliminary Be candidate sample is discussed in Section~\ref{sec:Be}. In Section~\ref{sec:RVs}, we discuss the membership of identified CBe 
stars in known open clusters and present the analysis of identifying runaway CBe stars in the sample based on their kinematics. We summarize our work in Section~\ref{sec:conc}.    

\section{LAMOST MRS optical Spectroscopy} \label{sec:observations} 
We obtained the optical spectra from the LAMOST at the Xinlong station of the National Astronomical Observatory. The LAMOST, also known as the Guoshoujing Telescope, is a 4 m quasi-meridian reflecting Schmidt telescope with an installment of 4,000 fibers within the 5 $\deg$ of Field of View. Starting in 2017, in addition to the existing low-resolution spectrographs ($R=1,800$), new medium-resolution spectrographs with a resolving power of $R=7,500$ were added to the telescope. LAMOST began the new five-year Medium Resolution Survey (MRS) in 2018 October. Part of the observing strategy for this survey is monitoring stars through multi-epoch observations. The LAMOST-MRS spectra are made with blue and red cameras. The spectra cover a wavelength range of 4950$-$5350 \AA\ for the blue arm and 6300$-6800$ \AA\ for the red arm \citep{liu2020}. In order to identify Be stars from the current released MRS database, we downloaded a total collection of 2,260,387 spectra with a signal-to-noise (S$/$N) greater than 10 per pixel in both blue and red bands from the LAMOST Data Release 7\footnote{\url{http://dr7.lamost.org/}}. The targets are brighter than the limiting magnitude of $Gaia\ G = 14.0$ mag for the selected spectral S/N criterion. The observations were made on nights between 2018 October and 2019 June. 

\begin{figure*}[!ht]
\plotone{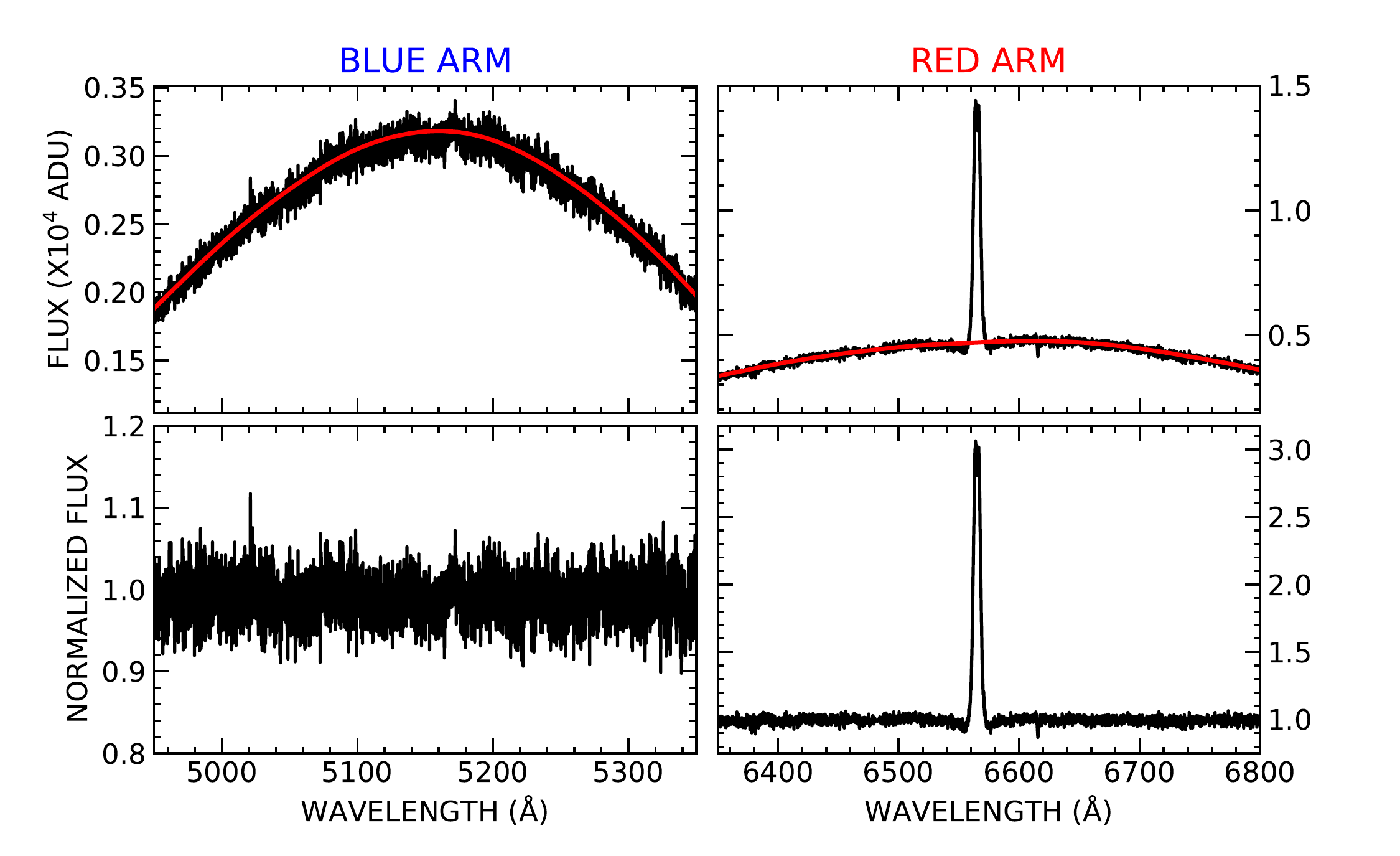}
\caption{The spectra for Be star HD 250854 (with LAMOST designation of J060334.97+221938.6) observed in both blue and red arms from LAMOST are 
shown in black on the top panels. The spline fits for the spectrum recorded in each arm are shown in red. The rectified spectrum for each band is shown in 
the bottom panel and has a continuum value about unity.}
\label{fig:Be_norm}
\end{figure*}

The spectra were reduced and the wavelength calibration was performed following the standard LAMOST 2D pipeline \citep{luo2015,ren2021}. Because cosmic rays were not fully removed by the pipeline, we adopted the Python package {\tt lapsec}\footnote{\url{ https://github.com/hypergravity/laspec}} from \citet{zhang2021} to remove any residual cosmic ray features present in the spectra. In order to rectify the observed spectra, we then transformed the spectra of the individual stars onto a uniform wavelength grid and rectified the spectra through a spline fitting to the selected pseudo continuum regions. In Figure~\ref{fig:Be_norm}, we show the spectra of a known Be star J060334.97+221938.6 (HD 250854) observed in both the blue and red bands (black) on the top panels, and the spline fits for each spectrum are shown in red. The rectified spectrum from each arm is shown on the bottom panels and now has a continuum value about unity. The final working product is a matrix of normalized flux, wavelength grid, and the observational dates recorded in the unit of the Modified Julian Date (MJD). 

\section{Selection of early-type stars from the MRS database} \label{sec:elimination}
\begin{figure*}[!ht]
\plotone{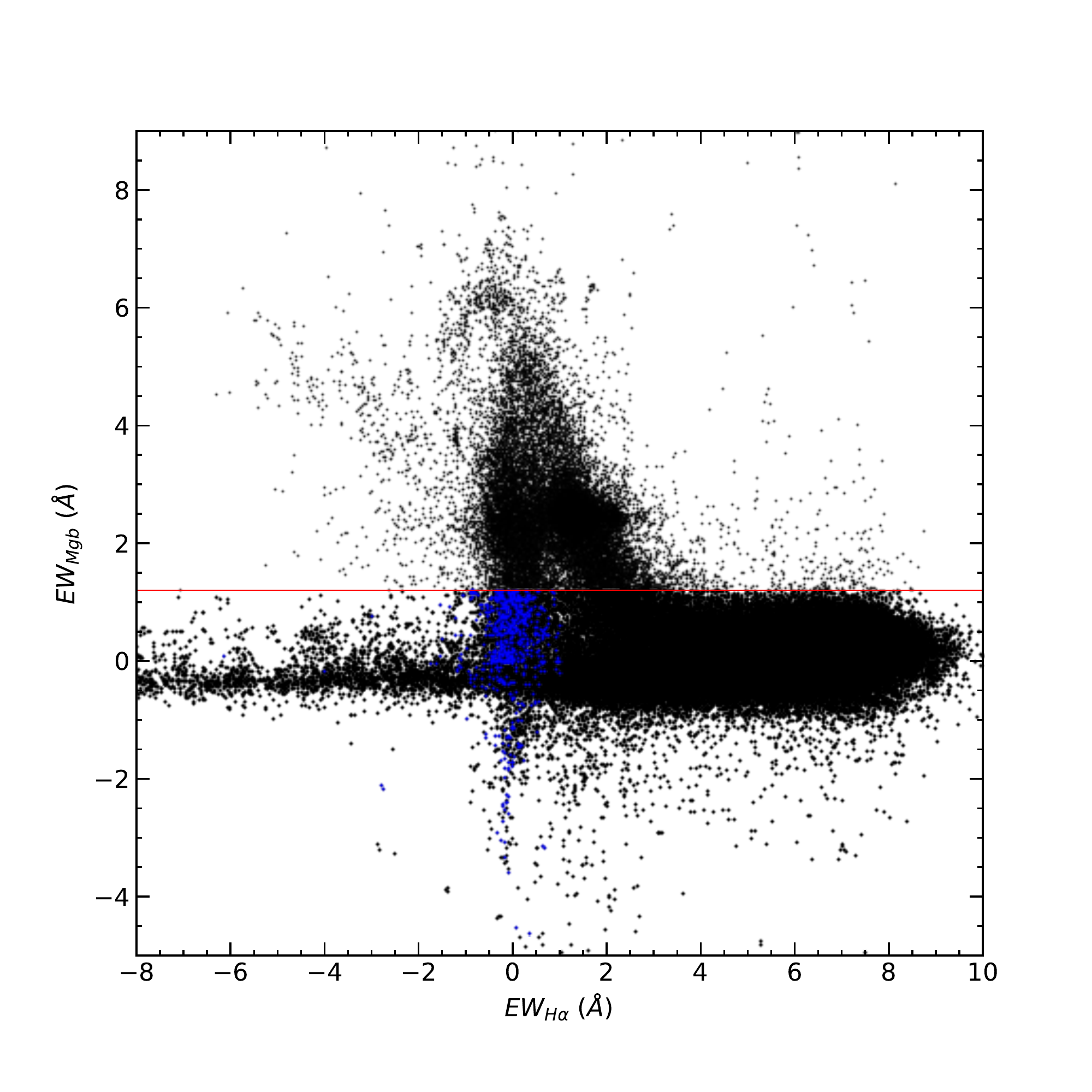}
\caption{The distribution of measured $EW$ values for 286,501 spectra from H$\alpha\ \lambda6565$ and \ion{Mg}{1\ $b$} triplet ($\sim \lambda5200)$ are
shown in black cross. A sample of 261,647 early-type stellar spectra were retained by keeping stars with measured $EW < 10.0$ for H$\alpha\ \lambda6565$ 
and $EW < 1.2$ for \ion{Mg}{1\ $b$} triplet (the black cross under the red horizontal solid line). By applying a reddening-free parameter from \citet{Comeron2005} using observations obtained from the 2\emph{MASS} all-sky survey, we further rejected 3,787 K- and M-type stars in the sample (blue cross). Our final working sample narrows down to a collection of 257,860 early-type stellar spectra. }
\label{fig:sample}
\end{figure*}

The archived spectra from the LAMOST MRS library comprise a broad range of stellar objects. The major goal of this work is to identify CBe stars from the 
database. We selected the spectra of early-type stars from the archived sample following the procedures described in \citet{Guo2021}. We first 
adopted the estimated effective temperatures ($T_{\rm eff}$) as given by the LAMOST stellar parameter pipeline (\emph{LASP}, \citealt{Wu2014}) to 
coarsely eliminate any late A-type to late F-type stars with 6,500 K $\le T_{\rm eff} \le$ 8,000 K from the MRS sample. This procedure narrows our initial 
working sample of 2,260,387 spectra down to 286,501. In order to remove any hidden late-type stars in the sample, we then adopted line profiles that are 
sensitive to the effective temperature, including H$\alpha\ \lambda6565$ and \ion{Mg}{1\ $b$} triplet ($\sim \lambda5200)$ series to compute their 
associated equivalent widths ($EW$) to form a distribution of $EW$ over the selected line indices. The $EW$ values for H$\alpha\ \lambda$ profile were 
measured over a wavelength range of $6548\sim6578$ \AA\ \citep{Cohen1998} and a bandpass of $5160\sim5193$ \AA\ 
was selected for \ion{Mg}{1\ $b$} triplet ($\sim \lambda5200)$ series \citep{Worthey1994}. In Figure~\ref{fig:sample}, we show the distribution of measured 
$EW$ values for both line profiles in the black cross. Metallic profiles develop their line strength towards late-type stars \citep{Gray2009}, we thus retained a 
sample of 261,647 early-type stellar spectra by keeping stars with measured $EW < 10.0$ for H$\alpha\ \lambda6565$ and $EW < 1.2$ for
\ion{Mg}{1\ $b$} triplet (the black cross shown on Figure~\ref{fig:sample} under the red horizontal solid line). Because stars with estimated 
$T_{\rm eff} < 3,100$ K are not given by the \emph{LASP} \citep{wu2011,luo2015}, we further excluded any K- and M-type stars were appearing in the sample from 
their position in a color-color magnitude plot using the IR photometric observations. \citet{Comeron2002} utilized the $J$-, $H-$, and $K_s$ photometric 
measurements for stars from the $2MASS$ all-sky survey to create an intrinsic color-color diagram to separate early-type and late-type 
stars in the Cygnus OB2 association (see their Figure 2), they further defined a reddening-free parameter of $Q = (J-H)-1.70(H-K_s)$ and reported that early-type stars have
typical $Q$ values about 0, while $Q = 0.4-0.5$ for K- and M-type stars \citep{Comeron2005,negueruela2007}. We cross-matched the stars in the sample 
with the $2MASS$ database \citep{cutri2003} using a circular searching aperture of 3$''$ to collect their photometric measurements, and we calculated the 
associated $Q$ values. By adopting the $Q$ criterion, we rejected 3,787 spectra for any K- and M-type stars appearing in the sample (shown as the blue cross in Figure~\ref{fig:sample}). The working sample is a collection of 257,860 early-type stellar spectra.    

\section{Identification of H$\alpha$ emission stars from LAMOST MRS spectra} \label{sec:identification_Be}
A broad range of astrophysical objects displays H$\alpha$ emission features in their spectra. Examples include planetary nebulae objects, in which the
emission feature originates from the ionized hydrogen gas at the ending stage of the stellar evolution of low-mass stars. Emission is also present in the 
spectra of pre-main-sequence young stellar objects, such as Herbig Ae/Be stars. Early-type main-sequence emission line stars, such as Oe, Be, or Ae 
stars, display H$\alpha$ emission features that are likely formed from the stellar wind or a decretion disk. In order to identify Be stars from the MRS 
database, we initiated a deep convolutional neural network approach to form a preliminary list of Be candidate stars from the identified early-type stars as 
described in Section~\ref{sec:elimination}.

\subsection{Preliminary selection of Be stars from the {\tt ResNet}} \label{subsec:RESNETS}
In recent years, machine learning techniques have been widely used to classify different types of astronomical images and signals based upon an automatic representation learning approach, such as the traditional kernel methods and deep Convolutional Neural Network (CNN) learning input data through the usage of multiple layers. The classification task is achieved by minimizing the generalization errors for the learning process, and the learning techniques differ in performance by their convergence rate to perform the recognition task.  

As described in \citet{lecun2015}, 
layers are applied to given input data to learn the morphological features from the data themselves. The output from each layer is evaluated by weighting 
factors to score the performance of the learning module from the desired recognition score, and subsequent layers learn from the input to identify classes 
in such a way to minimize the recognition errors. Outputs from the layers are combined to assemble the final representation of the identity. This technique is 
particularly efficient in identifying observational images with arrays of pixels in the modern large data era.

The representation of an identity learned from CNN benefits from gaining higher accuracy by adding more layers to the neural network architecture. Simply 
stacking more layers to the architecture may encounter a degradation issue, in which the accuracy of the recognition task reaches saturation and starts to 
decline \citep{he2016a}. We can reduce the computing time budget of applying CNN to a large set of images by combining the residual learning module with 
shortcut connections (the {\tt ResNet} algorithm), in which a layer or multiple layers are skipped to proceed with the identity mapping (see Figure 2 in \citealt{he2016b}). A detailed 
description of the building blocks and connection skipping are discussed in \citet{he2016a}.

Compared to the traditional image classification approach, such as the kernel method, the {\tt ResNet} algorithm achieves higher accuracy and maintains a faster converging rate. \citet{Allen2019} investigates the performance of these image classification schemes and provides provable evidence showing that the {\tt ResNet} reduces the complexity of training sample through the use of the multiple layers compared to the traditional ``one-shot" learning algorithm, and thus improves the classification efficiency. Also, smaller generalization errors were obtained from the {\tt ResNet} algorithms. For the same input data, traditional kernel methods can reach classification accuracy of 77\%$-$85\%, while the {\tt ResNet} achieves an accuracy of 96\% \citep{Recht2018, Arora2019}.

The {\tt ResNet} algorithm has been successfully applied to many astronomical image identification tasks. \citet{lir2019} applied the {\tt ResNet} module to 
search for potential galaxy Ly$\alpha$ emitter gravitational lens candidates from the SDSS \rom{3} spectroscopy survey. The {\tt ResNet} code was used to 
select pulsar candidates from the billion-scale database of the radio survey from the Five-hundred-meter Aperture Spherical radio Telescope 
\citep{wangh2019}. \citet{zhu2019} adopted several CNN variant networks to classify 28,790 
galaxy images into five classes based upon the morphological features of the galaxies from the Galaxy Zoo 2 database, and they concluded that the 
variant {\tt ResNet} model achieved state-of-art classification performance among the selected networks. There is a large collection of H$\alpha$ emission 
stars among those from the LAMOST MRS DR7, so we adopted this deep convolutional residual network to identify Be candidate stars in the database.  

In this work, we used the 18-layer {\tt ResNet} to identify Be star candidate stars in the sample. The {\tt ResNet} learning module was implemented in 
{\it torchvision.models}\footnote{\url{https://pytorch.org/vision/stable/models.html}} and obtained from the deep learning library of {\it PyTorch 1.9.0} 
\citep{Paszke2019}. The loss function of cross-entropy loss was adopted for the training, and the Adam optimizer from \citet{Kingma2014} was used to 
optimize the training model. We adopted a dynamic learning rate scheduler with an initial learning rate of 0.001. To improve the performance of the model and 
reduce the training epochs, the weights we loaded to the module were pre-trained on the {\it ImageNet} \citep{Deng2009}.

\subsubsection{Identifying H$\alpha$ peak featured spectra}
\label{subsubsec:preprocessing}

\begin{figure*}[!ht]
\gridline{\fig{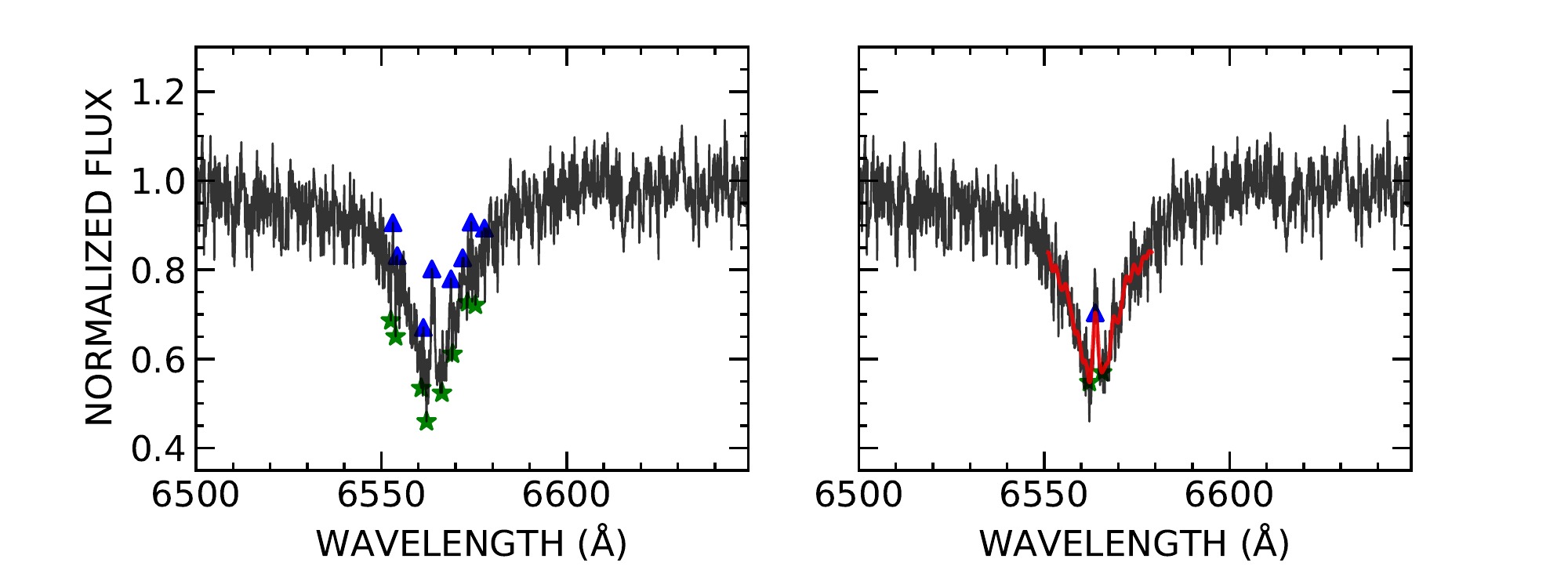}{1.0\textwidth}{(a) J023209.36$+$560939.5 }}
\gridline{
	\fig{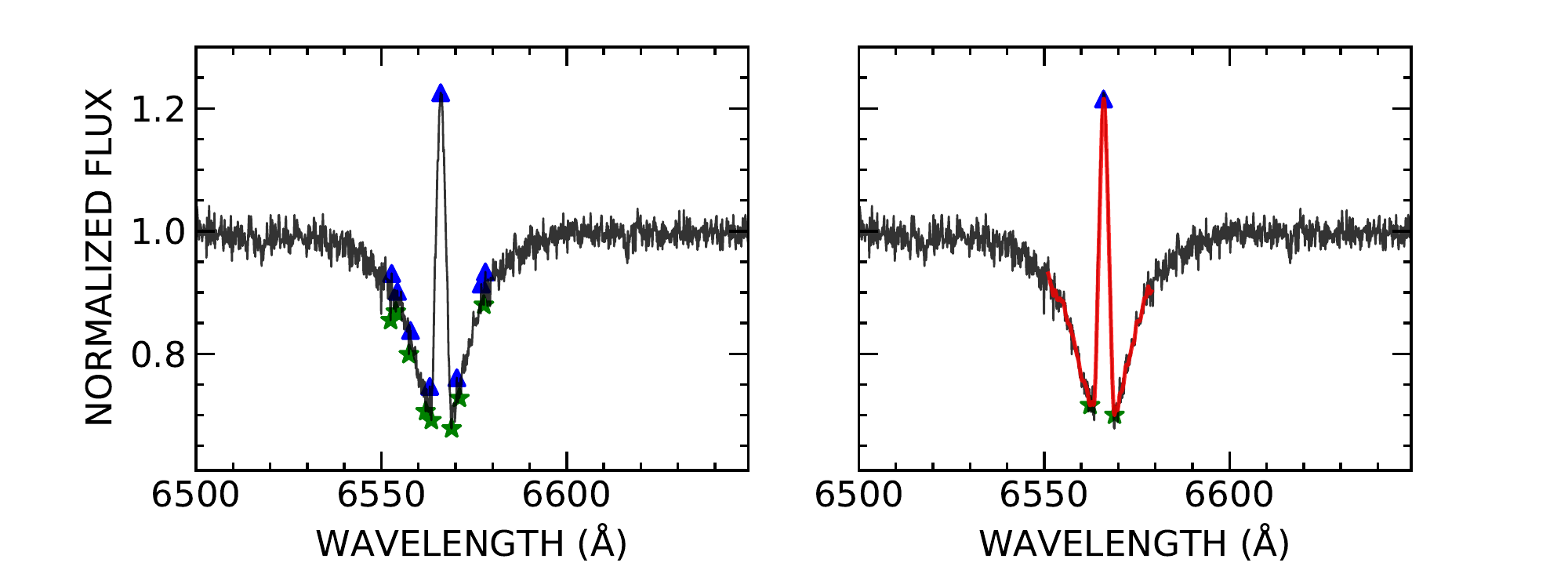}{1.0\textwidth}{(b) J071536.32$-$001018.3 } 
	}
\caption{The spectra of stars J023209.36$+$560939.5 (panel a) and J071536.32$-$001018.3 (panel b) were observed in the red band from the LAMOST MRS. \emph{Left panels}: Multiple locations of the valley pixels (blue star) and peak pixels (blue triangle) were determined from the H$\alpha$ profile over the wavelength region of $6551-6579$ \AA. \emph{Right panels}: By adopting a Bartlett window smoothing approach from \citet{Bartlett1950}, the location of the peak pixels with local maximum flux values is determined. Based on the S/N of the spectra, a window size of 15 pixels was chosen to smooth the profile over selected wavelength regions for J023209.36$+$560939.5, and the smoothed spectra are shown in red-line and over-plotted on the observed spectrum shown in black (top right panel). A window size of 9 pixels was selected in the case of J071536.32$-$001018.3 (bottom right panel).}
\label{fig:peak}
\end{figure*}

In order to apply the {\tt ResNet} to identify the Be spectra in the MRS database, we first need to construct the training sample. Both Be spectra, and non-Be spectra should be included to represent a broad range of spectral features in the sample. Because H$\alpha$ emission profiles feature the optical spectra of CBe stars, thus we are expected to obtain negative $EW$ values measured over the profile. However, in those cases of weak emission, as shown in Figure~\ref{fig:peak}, the H$\alpha$ peak is superimposed on the bottom of a broad absorption feature. The dominant absorption feature leads to positive calculated values of $EW$, and such stars will be missed from the selection using a simple criterion of positive $EW$ value.

In order to confirm the detection of the H$\alpha$ peak feature in the sample spectra, we restricted the profile within a wavelength region of $6551-6579$ \AA\ to search for the peak pixels with local maximum flux values. In Figure~\ref{fig:peak}, we show two example spectra of stars J023209.36$+$560939.5 and J071536.32$-$001018.3, observed in the red band from the LAMOST MRS in panels (a) and (b), respectively. As shown in the left panels of (a) and (b), multiple locations of the valley pixels (green star) and the peak pixels (blue triangle) were determined from the line profile. To avoid the ambiguity of locating the global maximum pixels from the local peak pixels, we adopted a Bartlett window approach from \citet{Bartlett1950} to smooth the spectra over the selected wavelength region. In the top right panel of Figure~\ref{fig:peak}, we experimented with various window sizes to smooth the selected portion of the H$\alpha$ profile for J023209.36$+$560939.5 to identify the peak feature, and 15 pixels were chosen for this spectrum with S/N $=13$. The smooth spectra over the selected region are shown in the red line and are over-plotted on top of the observed spectra shown in black. Due to the higher S/N for the spectrum of J071536.32$-$001018.3 (S/N $=50$), we selected a smaller window size of 9 pixels to perform the smoothing process (bottom right panel of Figure~\ref{fig:peak}). In Table~\ref{tab:Bartlett_window}, we list the S/N ranges and selected widths of the Bartlett window function for spectra in our sample.        

\begin{deluxetable*}{lc}
\tablenum{1}
\tablecaption{Selected width of the Bartlett window function \label{tab:Bartlett_window}}
\tablewidth{0pt}
\tablehead{
\colhead{S/N} & \colhead{Window width} \\
\colhead{Range} & \colhead{(pixel)}
}
\startdata
$\phn10 \le$ S/N $< 20$   	& 15	 \\
$\phn20 \le$ S/N $< 40$   	& 10  \\
$\phn40 \le$ S/N $< 60$   	& \phn9   \\
$\phn60 \le$ S/N $< 100$  	& \phn8 \\
$100 \le$ S/N $< 120$	& \phn7    \\
$120 \le$ S/N $< 150$ & \phn6    \\
 S/N $\ge$ 150       	& \phn5	 \\
\enddata
\end{deluxetable*}

\subsubsection{Constructing the training sample for the {\tt ResNet}}\label{subsubsec:training}
\begin{figure*}[!ht]
\plotone{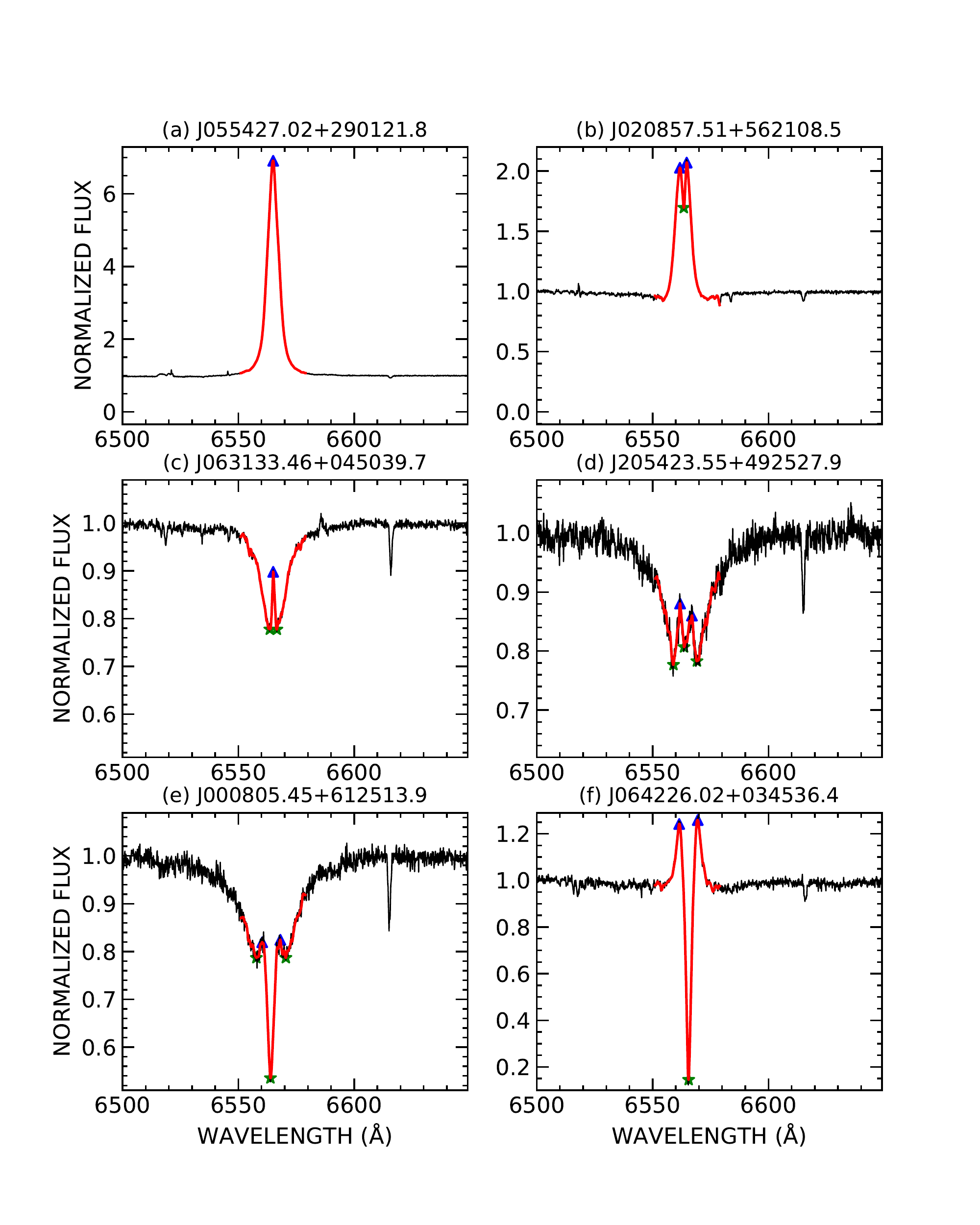}
\caption{A broad representation of the morphological features for the Be candidate spectra is selected for the training sample of {\tt ResNet}. The spectra included examples of single-peaked (panel a) and double-peaked emission profiles (panel b); emission profiles superimposed on the bottom of broad absorption profiles (panels c and d); and shell line profiles (panels e and f). The valley pixels, peak pixels, and the smoothed portion of the profile are labeled on the plot in each panel in the same format as Figure~\ref{fig:peak}. }
\label{fig:training_Be}
\end{figure*}

\begin{figure*}[!ht]
\gridline{\fig{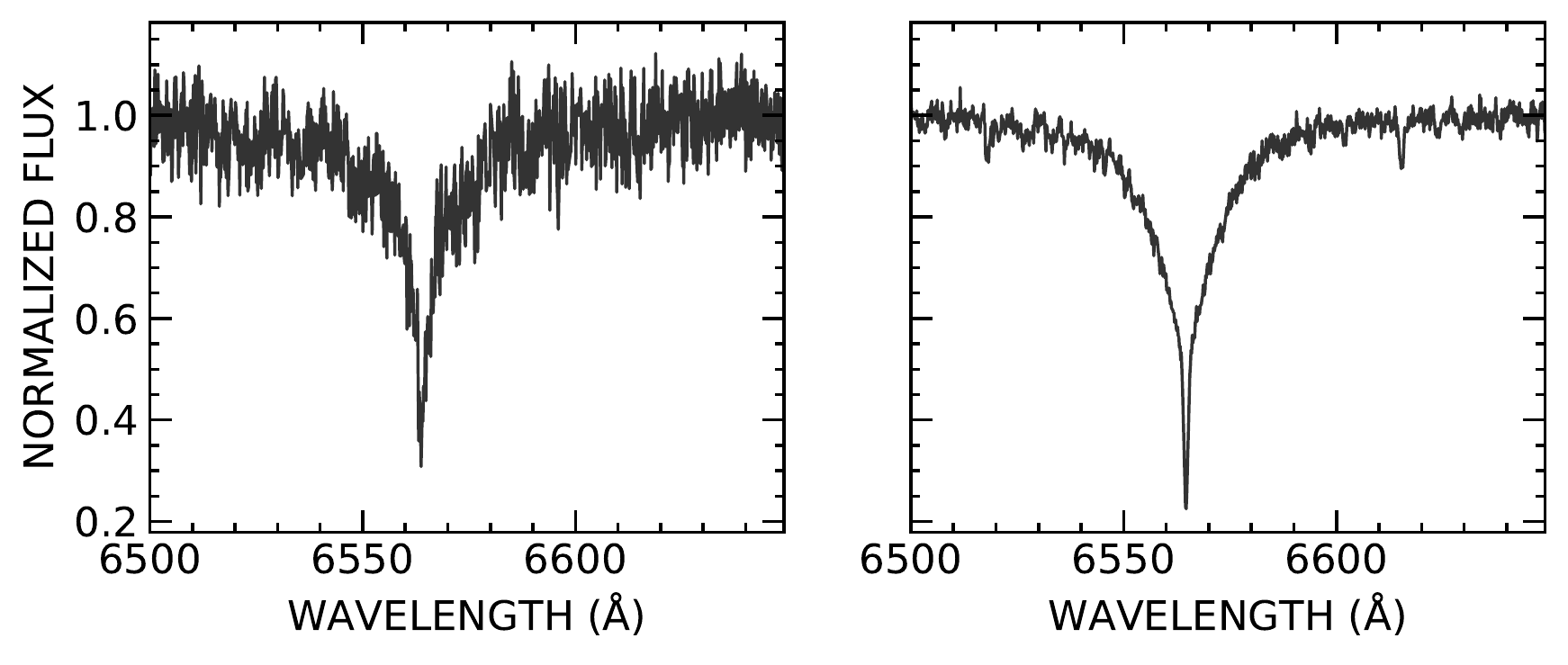}{1.0\textwidth}{(a) J000357.80+564258.4 (left) and J010743.93+585138.9 (right) }}
\gridline{\fig{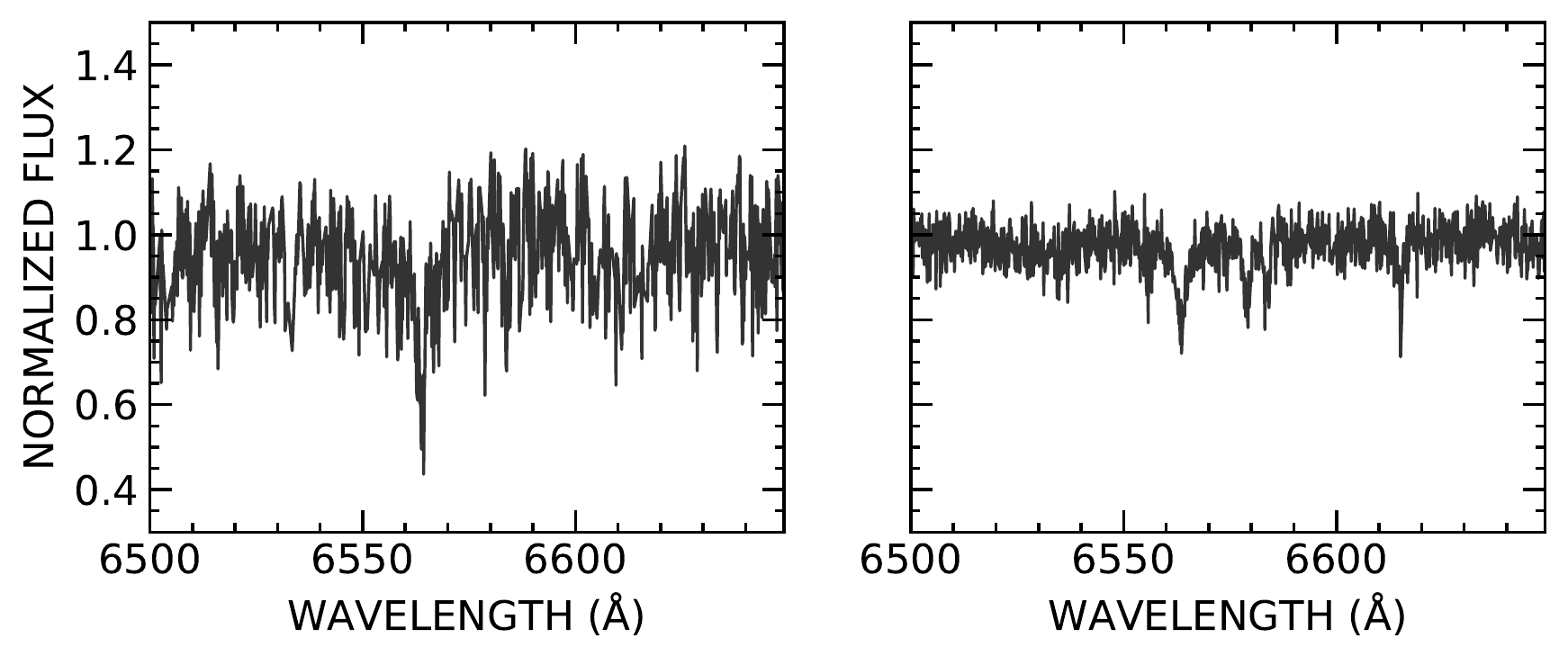}{1.0\textwidth}{(b) J000222.70+625403.1 (left) and J004953.72+643816.5 (right) }}
\caption{Example spectra for non-Be stars with H$\alpha$ absorption profiles randomly selected from the sample of 257,860 as discussed in Section~\ref{sec:elimination}. }
\label{fig:training_nonBe}
\end{figure*}

We visually inspected the spectra with identified H$\alpha$ features as discussed in Section~\ref{subsubsec:preprocessing} and found 1,042 spectra as the Be candidate spectra for the training sample. A broad representation of the morphological features for the Be candidate spectra is selected. In Figure~\ref{fig:training_Be}, we show example Be spectra displaying emission profiles, including both single peaked H$\alpha$ profile (panel a) and double emission peaks (panel b); those cases of weak emission profiles that are superimposed on the bottom of a broad absorption feature (panels c and d); and also absorption shell line profiles (panels e and f). In Table~\ref{tab:training_Be}, we list the LAMOST designation ID, the equatorial coordinates, the observational date, and the S/N for the found 1,042 Be candidate star. Since the training sample should include spectral features for both Be and non-Be spectra and maintain a symmetric number of observations for each category, we also visually selected 1,042 non-Be spectra from the sample of 257,860 early-type stellar spectra as discussed in Section~\ref{sec:elimination} to include any non-emission features in the training sample. Example spectra for the non-Be stars, such as strong H$\alpha$ absorption profiles (top panel) and weak featured spectra (bottom panel), are shown in Figure~\ref{fig:training_nonBe}. In Table~\ref{tab:training_nonBe}, we list the selected non-Be spectra in the same format as Table~\ref{tab:training_Be}.

We solely utilized the spectra observed in the red band of LAMOST MRS to perform the Be star identification task using the {\tt ResNet} algorithm.
The final training sample consists of 1,042 spectra of Be candidate stars and 1,042 spectral of non-Be stars. We generated their spectral images in a short wavelength range of $6530-6590$ \AA\ and maintained the same size of $256\times256$ pixels for each spectral image. We divided the sample into a training set comprising 80$\%$ of the spectral images and a validation set containing 20$\%$ of the spectral images.

\begin{deluxetable*}{ccccc}
\tablecaption{Be candidate spectra selected for the training sample \label{tab:training_Be}}
\tablewidth{0pt}
\tablehead{
\colhead{LAMOST spectrum} & \colhead{RA} & \colhead{DEC} & \colhead{Date} & \colhead{S/N} \\
\colhead{ID} & \colhead{(deg)} & \colhead{(deg)} & \colhead{(MJD)} & \colhead{}
}
\startdata
611012036 &	0.0161 & 63.9082 & 	58089.8069 & 	\phn63 \\
596908120 &	0.0878 & 55.4570 & 	58057.9354 & 	228  \\
595809135 &	0.1068 & 62.6628 & 	58055.8813 & 	\phn49 \\ 
611109135 &	0.1068 & 62.6628 & 	58089.8514 & 	\phn70 \\ 
596908134 &	0.1744 & 55.7225 & 	58057.9354 & 	145 \\ 
611012151 & 0.3529 & 63.5044 & 	58089.8069 & 	159 \\ 
609212029 &	0.6556 & 58.4759 & 	58087.7910 & 	145 \\ 
596906238 &	0.9008 & 55.8183 & 	58057.9354 & 	129 \\ 
611006213 &	1.1282 & 61.2467 & 	58089.8278 & 	\phn54 \\ 
611106196 &	1.1745 & 61.4988 & 	58089.8514 & 	\phn69 \\ 
\enddata
\tablecomments{This table is available in its entirety in machine-readable form. The first ten entries are shown here for guidance regarding its format and content.}
\end{deluxetable*}

\begin{deluxetable*}{ccccc}
\tablecaption{non-Be candidate spectra selected for the training sample \label{tab:training_nonBe}}
\tablewidth{0pt}
\tablehead{
\colhead{LAMOST spectrum} & \colhead{RA} & \colhead{DEC} & \colhead{Date} & \colhead{S/N} \\
\colhead{ID} & \colhead{(deg)} & \colhead{(deg)} &  \colhead{(MJD)} & \colhead{}
} 
\startdata
608405246 &	13.0407 & 	59.1671 & 	58085.8076 & 	\phn76 \\
684203235 &	13.4165 & 	37.2534 & 	58419.9174 & 	\phn98 \\
611804047 &	13.4377 & 	60.4097 & 	58090.8479 & 	\phn82 \\
608405084 &	13.5333 & 	58.7219 & 	58085.8076 & 	103 \\
611805039 &	13.7602 & 	59.0381 & 	58090.8479 & 	\phn84 \\
608404219 &	14.3077 & 	60.5316 & 	58085.8076 & 	110 \\
608404118 &	14.5391 & 	59.8807  &	58085.8076 &	105 \\
611809107 &	15.0439 & 	61.1058 & 	58090.8479 & 	\phn71 \\
684208174 &	15.6540 & 	37.0037 & 	58419.9174 & 	102 \\
611808033 &	15.6748 & 	59.8273 & 	58090.8479 & 	\phn69 \\
\enddata
\tablecomments{This table is available in its entirety in machine-readable form. The first ten entries are shown here for guidance regarding its format and content.}
\end{deluxetable*}

\subsubsection{Data augmentation}\label{subsubsec:augmentation}
The network achieves a higher performance of identifying input data if both the training errors and validation errors show a decreasing trend towards an increasing training time applied to the module (see Figure 1 of \citealt{shorten2019}). This can be achieved using a very powerful technique known as data augmentation, which is the standard procedure for training models in computer vision tasks. Such a process expands the dataset by generating different versions of the data with some transformations. Instead of simply increasing the number of samples, data augmentation can improve the performance of the model since it is a regularisation method that reduces the structural risk of the model and helps to avoid over-fitting \citep{8388338}.

We thus applied a series of geometric transformations to augment the diversity of the training set such that a broad representation of H$\alpha$ profile variations is included, and meanwhile, to preserve the morphological features of H$\alpha$ profiles. In this data augmentation process, the following techniques were applied to the spectra images: rotating the images slightly at random in the range of $-10^\circ$ to $+10^\circ$, zooming the images with multiples between 1.0 and 1.5, warping images in perspective, and flipping the images along the vertical axis. Data augmentation allows the network to learn these invariants in the dataset. It does not change the sample distribution and main characteristics of the data but creates variability and increases the diversity \citep{7797091,shorten2019}. As a result, the robustness and generalization ability of the model can be improved.

\subsubsection{Applying the {\tt ResNet}}\label{subsubsec:apply_ResNet}
\begin{figure}[ht!]
\plotone{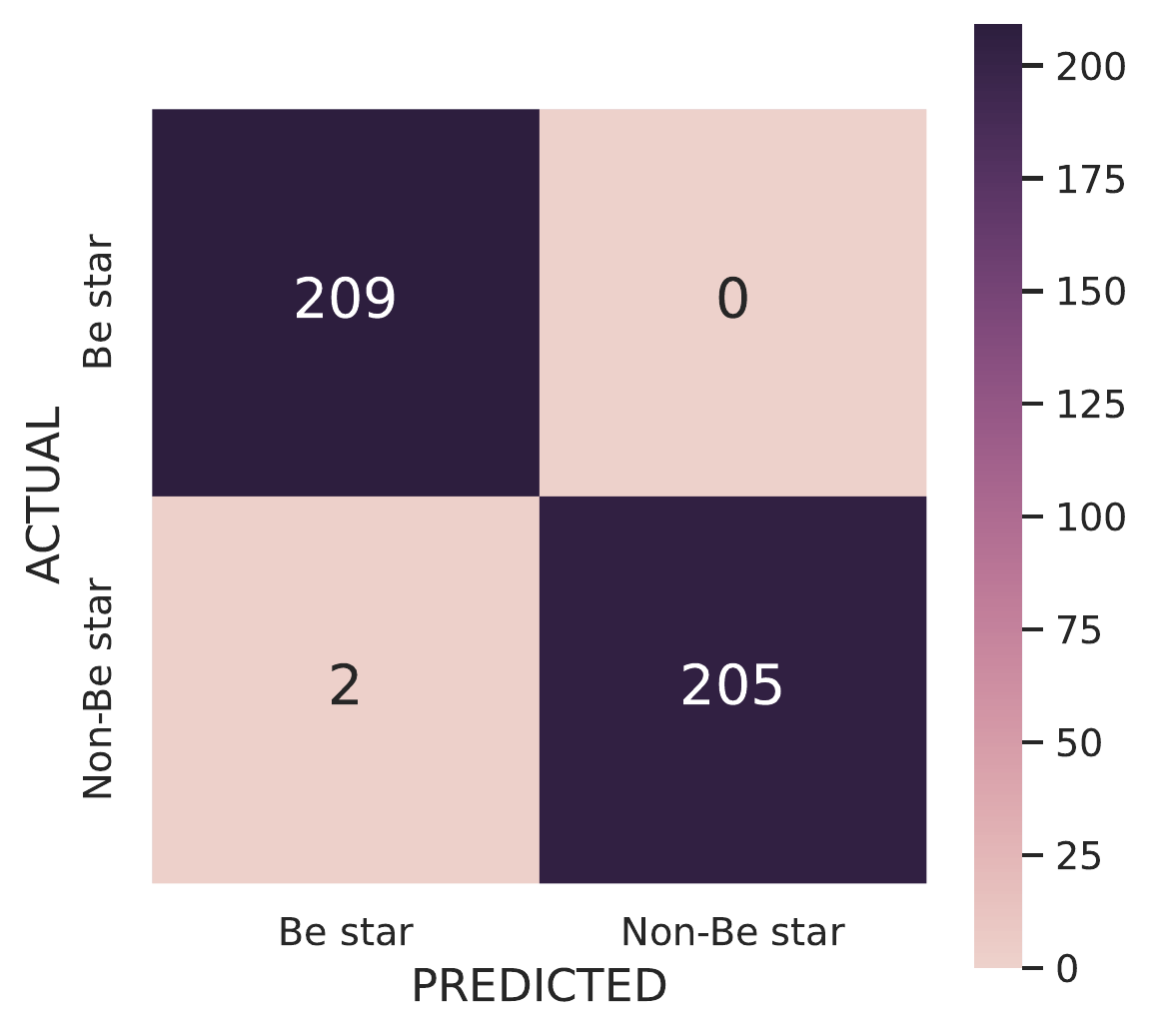}
\caption{Representation for the performance of the {\tt ResNet} applied to the H$\alpha$ featured spectra in the validation set. The upper left panel shows 
the TP detection of the learning module, and 209 spectra in the validation set with predicted Be spectra features were identified as the true label. The lower 
right panel shows the TN detection for 205 pre-labeled non-Be stars. The upper right panel shows the null detection of any pre-label non-Be stars identified 
as a Be star in the set, suggesting that none of the Be star spectral features were missed from the learning model. The lower left panel shows the spurious 
detection of Be stars due to misclassification. The {\tt ResNet} model achieves an accuracy of 99.5$\%$.} 
\label{fig:confusion_matrix}
\end{figure}

By applying the {\tt ResNet} to our training sample, the model learns a broad representation of H$\alpha$ features. We used accuracy as the performance metric to evaluate the quality of the 
model. The accuracy is defined as the ratio of the number of correct samples to the number of total samples. In 
Figure~\ref{fig:confusion_matrix}, we show the result of the identification task performed by the learning module for input spectra from the validation set 
after training the {\tt ResNet} for ten epochs. Among the set, 209 images with the pre-labeled spectral feature of Be stars were identified as True 
Positive (TP) detections (shown in the upper left panel of Figure~\ref{fig:confusion_matrix}), while 205 pre-labeled non-Be featured images were 
classified as True Negative (TN) by the {\tt ResNet} (lower right panel). None of the predicted non-Be spectral featured images were identified as a 
positive detection by the classifier, denoted as False Negative (FN), suggesting that all the Be star spectra features were recognized by the classifier (shown 
in the upper right panel of Figure~\ref{fig:confusion_matrix}). Such findings indicate that our model achieves an accuracy of 99.5$\%$, and most of the sample 
spectra can be classified well by the module. However, two images with predicted labels of Be spectral features were labeled as non-Be type stars 
from the learning module (lower left panel of Figure~\ref{fig:confusion_matrix}), indicating that the identified Be star sample may include a small fraction of 
spurious detection due to the misclassification (denoted as False Positive, FP). We thus visually inspected the sample of identified Be candidate stars 
obtained from the {\tt ResNet} and arrived with a preliminary sample of 1,134 
Be candidate stars. 

\subsection{The morphological classification of the H$\alpha$ emission stars}
\begin{figure*}[ht!]
\plotone{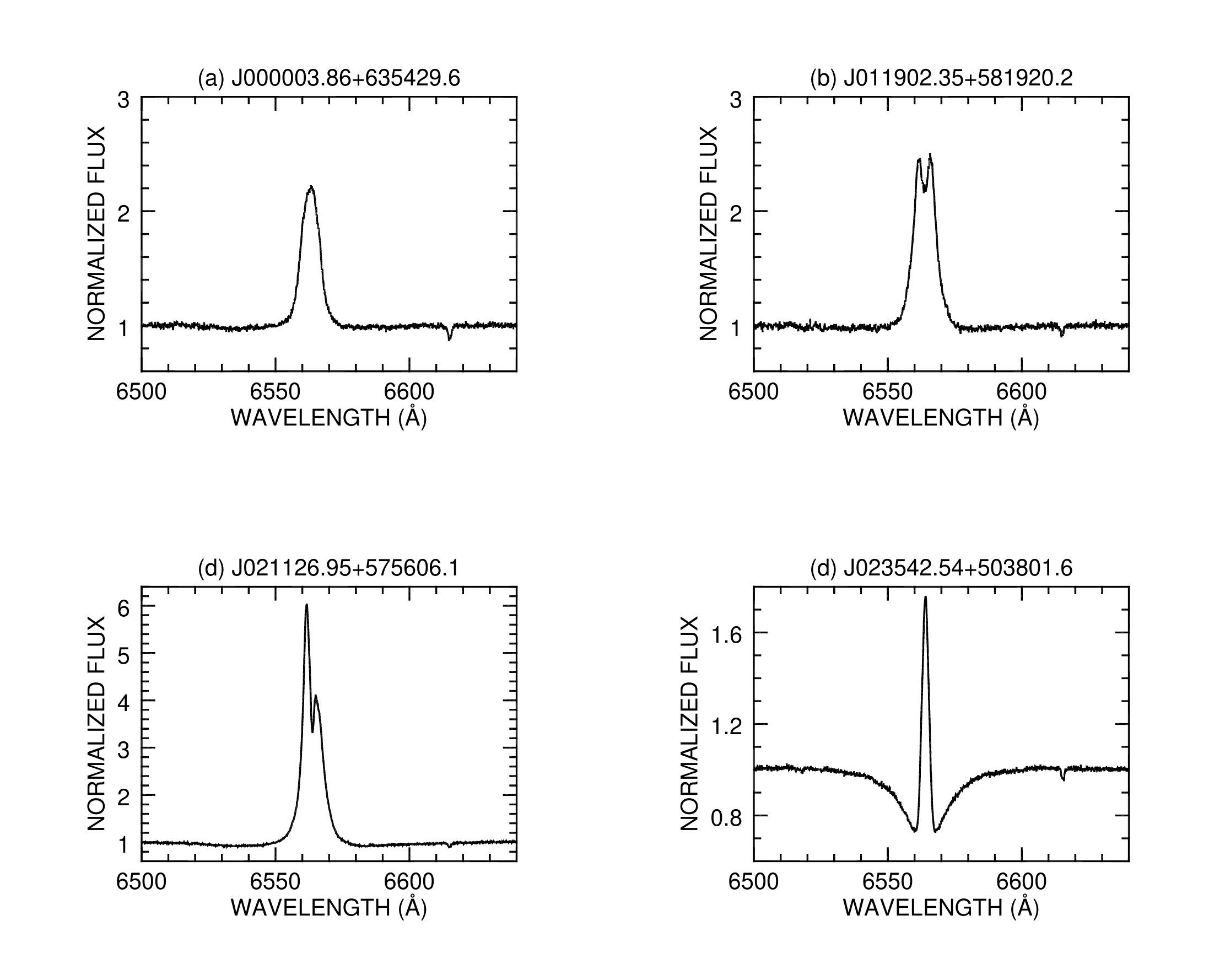}
\caption{The H$\alpha$ emission profiles of stars selected from the LAMOST MRS DR7 database with various morphological features. (a) 
J000003.86+635429.6 with a single emission peak, (b) J011902.35+581920.2 with symmetric double emission peaks, (c) J021126.95+575606.1, 
and (d) J023542.54+503801.6 with a narrow emission peak superimposed on top of an absorption profile.}
\label{fig:emission}
\end{figure*}

Based upon the morphological features of the H$\alpha$ profile shown in the red spectra of the Be candidate stars identified from the usage of {\tt ResNet}, 
we visually inspected the spectra of the stars and grouped them into two categories of emission lines and shell line features. Among the sample of 
1,134 Be candidate stars, 822 stars display H$\alpha$ emission profiles in their spectra, 113 stars show absorption shell line features superimposed 
on top of the broad H$\alpha$ emission profiles, and 199 stars display shell line features superimposed on top of the absorption H$\alpha$ profile. In 
Figure~\ref{fig:emission}, we show examples of profiles with single H$\alpha$ emission (panel a), symmetric double emission peaks 
(panel b), asymmetric double emission peaks (panel c), and a narrow emission superimposed on top of an absorption profile (panel d). 
We also classified the H$\alpha$ shell line featured stars into subgroups based upon their shell line width and symmetry, and the complete catalogs listing 
the morphological classification of these spectra are discussed in the Appendix.    

\placefigure{fig:emission}

\section{Identification of classical \texorpdfstring{B\MakeLowercase{e}}{Be} stars from the preliminary sample} \label{sec:Be}
\subsection{Literature cross-matching}
Following the selection of the preliminary sample of Be stars reported in Section~\ref{sec:identification_Be}, we applied a three-step test to the 
selected stars to identify the CBe stars. We first adopted the primary defining criterion of the CBe stars, the presence of the Balmer emission line 
feature in the spectra to visually select stars displaying H$\alpha$ emission using spectra observed from the LAMOST MRS red arm. 
This criterion resulted in a collection of 822 stars identified in the sample of 1,134 Be candidate stars. We then performed a literature search to cross-
match stars in the sample with prior classified CBe stars. The major resources include the catalog of 1,159 Be stars by \citet{jaschek1982}; 
the BeSS database \citep{neiner2011}, a total collection of 2,195 CBe stars as of 2021 April 13; 213 Be stars from the APOGEE 
survey by \citet{chojnowski2015}; detections of 192 CBe candidates stars from the LAMOST DR1 Low-Resolution Survey (LRS) from \citet{Lin2015}; a comprehensive compilation of 1,991 sources from the BeSS, APOGEE, and the INT 
Photometric H$\alpha$ Survey of the Northern Galactic Plane (IPHAS) from \citet{chen2016}; the Be catalog of 5,603 stars identified 
from the LAMOST DR2 database from \citet{hou2016}; the 693 newly classified CBe candidate stars found 
using {\it Gaia} DR2 observations by \citet{vioque2020}; and the 2,716 hot emission-line identified using the LAMOST DR5 LRS by
\citet{Shridharan2021}. We also cross-matched stars in our sample with those discussed in publications to search for any known 
Herbig Ae/Be stars, and 12 such stars were found from the literature searches. Some of the target stars may display H$\alpha$ emission being 
contaminated by ionized nebulae in \ion{H}{2} regions. Three stars in the sample were found from the cross-matching of known stars in \citet{hou2016}, who
listed a collection of 3,600 contaminated emission stars. The emission lines of H$\alpha$, [\ion{N}{2}] $\lambda$6548, $\lambda$6583, \ion{He}{1} $\lambda$6678,
and [\ion{S}{2}] $\lambda$6717, $\lambda6731$ are shown in the red spectra, and [\ion{O}{3}] $\lambda$4959, $\lambda5007$ feature is shown in the blue spectra
of these stars. The number of stars found from the literature search appears in Table \ref{tab:literature}. 
After removing duplications, a total collection of 151 CBe stars are found from the literature search, 
resulting in the preliminary sample narrowed to 671 new CBe candidate stars. We show the LAMOST designation ID, right ascension (RA), declination (DEC), star name, 
spectral classification, and the associated reference of the cross-matched known CBe stars in Table~\ref{tab:knownBe}.  

\begin{deluxetable*}{lcl}
\tablecaption{Common stars found from literature cross-matching \label{tab:literature}}
\tablewidth{0pt}
\tablehead{
\colhead{Object type} & \colhead{Number of stars found} & \colhead{Reference} 
}
\startdata
CBe stars & 43 & \citet{neiner2011}   \\ 
CBe stars & 10 & \citet{chojnowski2015}   \\ 
CBe stars & \phn3 & \citet{Lin2015} \\
CBe stars & \phn9 & \citet{chen2016}   \\ 
CBe stars & 63 & \citet{hou2016}   \\ 
CBe stars & 36 & \citet{vioque2020}   \\
Herbig Ae/Be stars & \phn3 & \citet{herbst1999}   \\ 
Herbig Ae/Be stars & \phn1 & \citet{vieira2003} \\
Herbig Ae/Be stars & \phn2 & \citet{neiner2011}   \\ 
Herbig Ae/Be stars & \phn1 & \citet{hou2016}   \\ 
Herbig Ae/Be stars & \phn5 & \citet{vioque2020}   \\ 
H\rom{2} regions & \phn3 & \citet{hou2016}   \\   
\enddata
\end{deluxetable*}

\subsection{Spectral type collection}
We then cross-matched the stars with the spectral classification catalogs of \citet{skiff2014} and \citet{kharchenko2009} to collect their spectral types.
We complement the search with the usage of SIMBAD \citep{wenger2000} for stars not found in these catalogs. In the case of
stars with contradicting spectral types, we adopted the classification from the most recent available publication. We conservatively 
excluded stars from the sample with no record of spectral type. Among the H$\alpha$ emission stars with archived spectral types from literature, we 
identified eight O-type emission line stars from the sample and listed their LAMOST designations, star names, equatorial coordinates, 
spectral types, and the associated references in Table~\ref{tab:Oe}. Stars with spectral types later than O8 
and earlier than A2-type only were selected as potential CBe candidate stars, and we also excluded stars with the luminosity class of 
\rom{1} and \rom{2} appearing in their spectral classification. A collection of 222 stars were cross-matched from the catalogs that
satisfy this second test.   
 
\begin{deluxetable*}{clcclc}
\tablecaption{Oe stars found from the LAMOST MRS DR7 sample \label{tab:Oe}}
\tablewidth{0pt}
\tablehead{
\colhead{LAMOST}  & \colhead{Star} & \colhead{RA}  & \colhead{DEC}  & \colhead{Spectral} & \colhead{Spectral Classification}  \\
\colhead{designation}  & \colhead{Name} &  \colhead{(deg)} & \colhead{(deg)} & \colhead{Classification} & \colhead{Reference}
}
\startdata
J000441.87+612955.7	&  EM VES 682		&  \phn\phn1.1745 	 &	61.4988 	& O9 \rom{3} &  \phn3 \\
J000925.08+624718.2 	&  EM GGR 172 	&  \phn\phn2.3545 	 &	62.7884 	& O9 &  \phn4 \\
J040320.74+511852.5 	&  BD+50 886 		&  \phn60.8365 &	51.3146 	& O4 \rom{5}((c)) &  85 \\ 
J052000.64+385443.4	&  LS V +38 12 	&  \phn80.0027 &	38.9121 	& O6.5 \rom{5}((f))  &  85 \\
J202937.57+405108.5	&  UCAC4 655-091969	&  307.4066 &	40.8524 	& O8 \rom{3}((f))  &  85 \\
J203457.84+414354.2 	&  RLP 1252		&  308.7410 &	41.7317 	& O8 \rom{5}e  &  85 \\
J222830.98+560054.7 	&  EM AS 491		&  337.1291 &  56.0152	& O5  &  \phn4 \\
J234528.61+631616.7	&  LS \rom{1} +62 14		&  356.3692 &	63.2713 	& O5: \rom{1}afpe  &  85 \\
\enddata
\tablecomments{Indices of references: (3) \citet{Martin1972} (4) \citet{Reed2003} (85) \citet{Li2021}. This table is available in machine-readable form.}
\end{deluxetable*}

\subsection{The IR test}
We caution that the candidate CBe stars in the sample may be contaminated by the presence of pre-main-sequence Herbig Ae/Be stars,
which also display H$\alpha$ emission profiles. The similarity of their H$\alpha$ emission features often confuses the 
classification of the CBe stars. 
\citet{finkenzeller1984} suggested that the Herbig Ae/Be stars display 
stronger IR excess compared to CBe stars, and the two types of stars can be distinguished from the IR photometry through position in a $(H-K, K-L)$ color-
color diagram. \citet{hou2016} demonstrated the applicability of replacing the $L$ band with the {\it WISE} $W_1$ band to separate the two classes 
in a $(H-K, K-W_1)$ color-color diagram. We thus adopt this approach to identify the CBe stars in our 
preliminary sample. We cross-matched stars in our sample with those in the \emph{AllWISE} Data Release from \citet{cutri2021}, and 206 
stars were found with recorded measurements in the $J$, $H$, $K$, and $W_1$ bands. Because of the presence of interstellar dust, 
extinction corrections were applied to the archived photometric measurements. We obtained the reddening values of the stars from the empirical 
three-dimensional dust map from \citet{green2019} based upon their equatorial coordinates and distance. The latter quantity was
collected from the \emph{Gaia} EDR3 \citep{bailer-jones2021}. We adopted the extinction ratio $A_\lambda/A_V$ values listed in Table 3 of 
\citet{cardelli1989}, which are based upon a ratio of total to selective extinction of $R_V = A_V/E(B-V) = 3.1$ \citep{fitzpatrick1999}. 
The adopted ratios have values of 0.282, 0.190, and 0.114 for the $J$, $H$, $K$ filters, respectively. We then adopted the extinction coefficient 
for the \emph{WISE} $W_1$ band from \citet{yuan2013} with $A_{W_1}/A_{V} = 0.061$. According to the selecting criterion given 
by \citet{hou2016}, Herbig Ae/Be stars are founded in the range of $H-K > 0.4$ and $K-W_1> 0.8$ in the color-color magnitude diagram, 
while the CBe stars have $H-K < 0.2$ and $K-W1 < 0.5$. We applied these criteria to the corrected photometric measurements to identify CBe 
and Herbig Ae/Be stars. This leads to a total identification of 183 new CBe stars and two Herbig Ae/Be stars. 
In Table~\ref{tab:newBe}, we list the LAMOST designation ID, star name, right ascension (RA), declination (DEC), spectral classification, 
reference of the classification, the \emph{AllWISE} photometric measurements of $J$, $H$, $K$, $W_1$ bands, their associated errors, 
and the reddening for each new detection.  

\begin{figure*}
\gridline{\fig{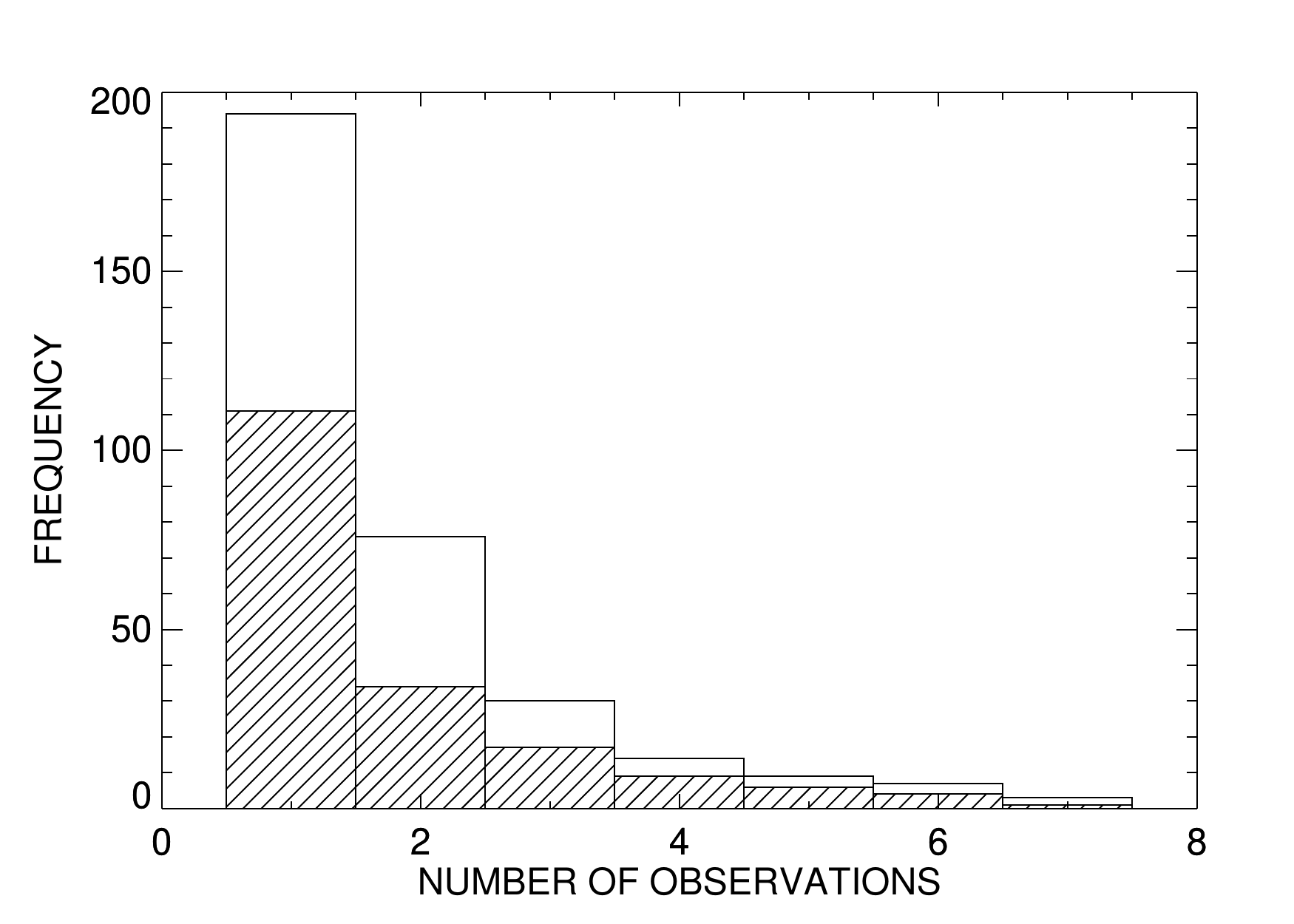}{0.5\textwidth}{(a) The distribution of the number of observations.}
	\fig{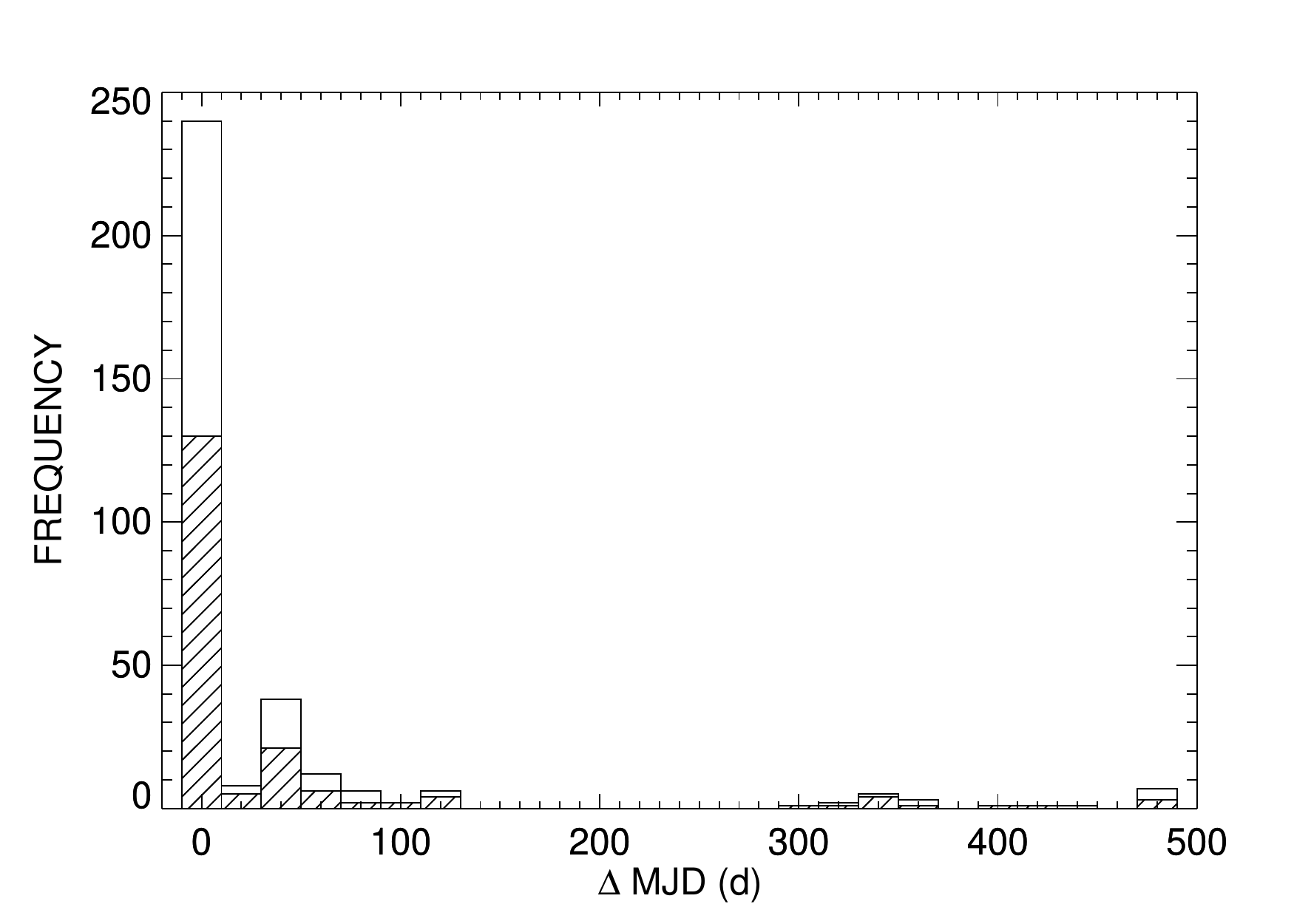}{0.5\textwidth}{(b) The distribution of the observational baseline. } 
	}
\caption{The left panel plots the distribution of the number of observations for the full sample of 334 CBe stars found from the LAMOST MRS DR7 database, and the 
right panel plots the distribution of the observational baseline ($\Delta$ MJD = MJD$_{max} - $ MJD$_{min}$) of the sample stars. The full sample is shown in solid line, 
and the new 183 detections are shown with filled lines. }
\label{fig:histogram}
\end{figure*}

Based on literature cross-matching and new identifications, a total collection of 334 CBe stars are found from the LAMOST MRS DR7 database. 
In Figure~\ref{fig:histogram}, we show the distribution with the number of observations of the entire sample (solid line) and the 183 new detections 
(line filled) of all the CBe stars in the left panel. One hundred ninety-five of the CBe stars were observed in a single visit. The right panel displays the 
number distribution of the CBe stars versus the observational baseline.     

\begin{deluxetable*}{clcclc}
\tablecaption{Known Be stars found from the LAMOST MRS DR7 sample \label{tab:knownBe}}
\tablewidth{0pt}
\tablehead{
\colhead{LAMOST}  & \colhead{Star} & \colhead{RA}  & \colhead{DEC}  & \colhead{Spectral} & \colhead{Spectral Classification}  \\
\colhead{designation}  & \colhead{Name} &  \colhead{(deg)} & \colhead{(deg)} & \colhead{Classification} & \colhead{Reference}
}
\startdata
J000124.69+633015.7 	&	BD+62 2346 	&	 \phn0.3529	& 	63.5044	&	B0 V		&       \phn3  \\
J000145.61+632758.0 	&        \nodata           &       \phn0.4401	&	63.4661	&      B		&	 \phn1 	\\
J010642.79+615941.9 	&        \nodata           &      16.6783		&	61.9950	&     \nodata	& 	\nodata \\        
J011833.07+582230.5 	&	HD 236689 	&	19.6378		& 	58.3751 	&	B1.5 Vep\_{sh} 	&	10   \\
J011854.11+581206.8 	&	BD+57 243 	&	19.7255		& 	58.2019	& 	B3         	&	66   \\
J012319.41+573851.7 	&	BD+56 259 	&	20.8309		& 	57.6477 	&	B3e       	& 	33   \\
J012401.03+585036.1 	&	VES 704 		&	21.0043 		&	58.8434 	&	Be         	&	67   \\
J012429.14+601723.2 	& 	BD+59 246 	&	21.1214		& 	60.2898 	&	Ae        	& 	68   \\
J012718.47+600653.6 	&	BD+59 250 	&	21.8270		& 	60.1149 	&	B0e        	&	19   \\
J012924.35+590557.3 	&	BD+58 247 	&	22.3515		& 	59.0993 	&	B1/2e      	&	69   \\
\enddata
\tablecomments{Indices of references: (1) \citet{Brodskaya1955} (3) \citet{Martin1972} (10) \citet{Morgan1955} (19) \citet{Merrill1942} (33) \citet{Merrill1949} (66) \citet{Hayford1932}  (67) \citet{Miller1951} (68) \citet{Merrill1950}  (69) \citet{McCuskey1974}. This table is available in its entirety in machine-readable form. The first ten entires are shown here for guidance regarding
its format and content.}
\end{deluxetable*}

\begin{deluxetable*}{clcclcccccccccc}
\rotate
\tabletypesize{\scriptsize}
\tablecaption{Newly identified CBe stars from the LAMOST MRS DR7 sample \label{tab:newBe}}
\tablewidth{0pt}
\tablehead{
\colhead{LAMOST}  & \colhead{Star}   & \colhead{RA}  & \colhead{DEC}  & \colhead{Spectral} & \colhead{Reference}  & \colhead{$J$} & \colhead{$\sigma_{J}$} & \colhead{$H$} &   \colhead{$\sigma_{H}$} & \colhead{$K$} &  \colhead{$\sigma_{K}$} 
& \colhead{$W_1$}   & \colhead{$\sigma_{W_1}$} & \colhead{$E(B-V)$} \\
\colhead{designation}  &  \colhead{Name} & \colhead{(deg)} & \colhead{(deg)} & \colhead{Classification} & \colhead{} & \colhead{(mag)}  & \colhead{(mag)} & \colhead{(mag)}  & \colhead{(mag)} & \colhead{(mag)}  & \colhead{(mag)}  & \colhead{(mag)}  & \colhead{(mag)}   & \colhead{(mag)}  
}
\startdata
J000035.31+623317.7 &TYC 4018-3553-1 &  \phn0.1472 &62.5549	 &	B2	&  \phn1 &11.030& 0.021& 10.944 &0.030& 10.921& 0.023& 10.591 &0.023 &0.320 \\
J000430.77+611448.2 &EM* VES 681   	&   \phn1.1282& 61.2467 	& 	B2 	&  \phn2 & \phn9.484& 0.022&  \phn9.274& 0.028&  \phn9.063 &0.022 & \phn9.082& 0.022& 0.560 \\
J000441.87+612955.7 &EM* VES 682   	&   \phn1.1745 &61.4988	 &	O9 \rom{3} &  \phn3& 10.667& 0.024& 10.527& 0.031& 10.448& 0.030& 10.426& 0.022& 0.392 \\
J000925.08+624718.2 &EM* GGR 172  	&   \phn2.3545 &62.7884	& 	O9	&  \phn4& 10.092& 0.018&  \phn9.857& 0.017&  \phn9.632& 0.022&  \phn9.274& 0.023& 0.436 \\
J001324.32+591313.4 &LS I +58 13   	&   \phn3.3514 &59.2204 &	OBe  & \phn5 & \phn9.407 &0.020 & \phn9.196& 0.020&  \phn9.022 &0.022&  \phn9.132 &0.023& 0.816\\ 
J001930.74+600218.4 &LS I +59 39    	&   \phn4.8781 &60.0385 &	OB  	&   \phn5 & \phn9.592 &0.022 & \phn9.390& 0.032&  \phn9.187& 0.022 & \phn9.054 &0.023& 0.570 \\
J002459.27+591931.4 &EM* GGR 183 	&   \phn6.2470 &59.3254 &	A  	&    14& 11.371& 0.028 &11.157& 0.033 &10.955 &0.028& 10.645 &0.023 &0.581 \\
J002629.20+600127.9 &LS I +59 42      	&   \phn6.6217 &60.0244 &	OBe &  \phn5& 10.048 &0.027&  \phn9.715 &0.032&  \phn9.430& 0.024  &\phn9.638 &0.024& 0.820\\ 
J002935.50+601817.8 &EM* GGR 187  	&   \phn7.3979 &60.3049 	&	Ae 	&   \phn6 &10.796 &0.019 &10.708& 0.027 &10.662& 0.025 &10.624 &0.023 &0.450 \\
J004315.45+595219.9 &EM* GGR 201  	&   10.8144 &59.8722 &	Ae 	&  \phn6 &10.748 &0.019& 10.527& 0.028& 10.289& 0.023& 10.017 &0.022& 0.520 \\ 
 \enddata
\tablecomments{Indices of references: (1) \citet{Brodskaya1955} (2) \citet{Brodskaya1953} (3) \citet{Martin1972} (4) \citet{Reed2003} (5) \citet{Hardorp1959} (6) \citet{Gonzalez1956} (14) \citet{wenger2000}. This table is available in its entirety in machine-readable form. The first ten entires are shown here for guidance regarding
its format and content.}
\end{deluxetable*}

\placefigure{fig:distribution}

\section{Open cluster membership and Kinematics of the \texorpdfstring{CB\MakeLowercase{e}}{CBe} stars} \label{sec:RVs}
We are interested in determining the potential membership of the identified 334 CBe stars in the known open clusters (OC).
We examined the OC data from the work of \citet{cantat2018}, who compiled the star members and cluster parameters for 1,229 OCs using 
{\it Gaia} DR2 astrometric data; a compilation of 226 OCs using the data from {\it Gaia} DR2, GALAH$+$, and APOGEE DR16 from 
\citet{spina2021}; and a comprehensive collection of fundamental parameters, including the mean radial velocity ($V_{r}$), distance, age, 
metallicity, and extinction for 1,743 OCs from \citet{dias2021}. We collected the equatorial coordinates ({\it RA} and {\it DEC}), the proper motion ($
\mu_\alpha$ and $\mu_\beta$), the parallax ($\pi$), and the associated uncertainty parameters for each of the CBe stars from the {\it Gaia} EDR3 
\citep{gaia2016,gaia2021} using a circular search aperture of 1$''$. A quality filter of the Renormalised Unit Weight Error ({\tt RUWE}) $> 1.4$ was 
applied to omit any problematic astrometric measurements.  

We first cross-matched the equatorial coordinates and proper motions of the CBe stars 
with member stars with a classified membership probability $> 60\%$ of the known OCs. Because stars may be located either in the foreground or 
the background of the OC, we then cross-matched the parallaxes of CBe stars with those measurements of cluster member stars. We identified 41 
CBe stars that are associated with known OCs, and report their LAMOST designation and cluster name in the first two columns of 
Table~\ref{tab:OC}, respectively. Fundamental parameters of cluster radial velcoity $V_{r}$, derived distance, cluster age, metallicity, extinction, and 
their associated errors taken from the compilation from \citet{dias2021} are listed in columns 3 to 12.  Among these identified cluster member stars, 
13 of them are located in the same cluster of NGC 884 ($\chi$ Persei), \citet{Marsh2012} conducted a spectroscopic investigation including 46 Be
stars in this cluster to determine their physical properties and reported that they are likely more evolved than stars found in similar open clusters. 

\begin{figure*}[ht!]
\plotone{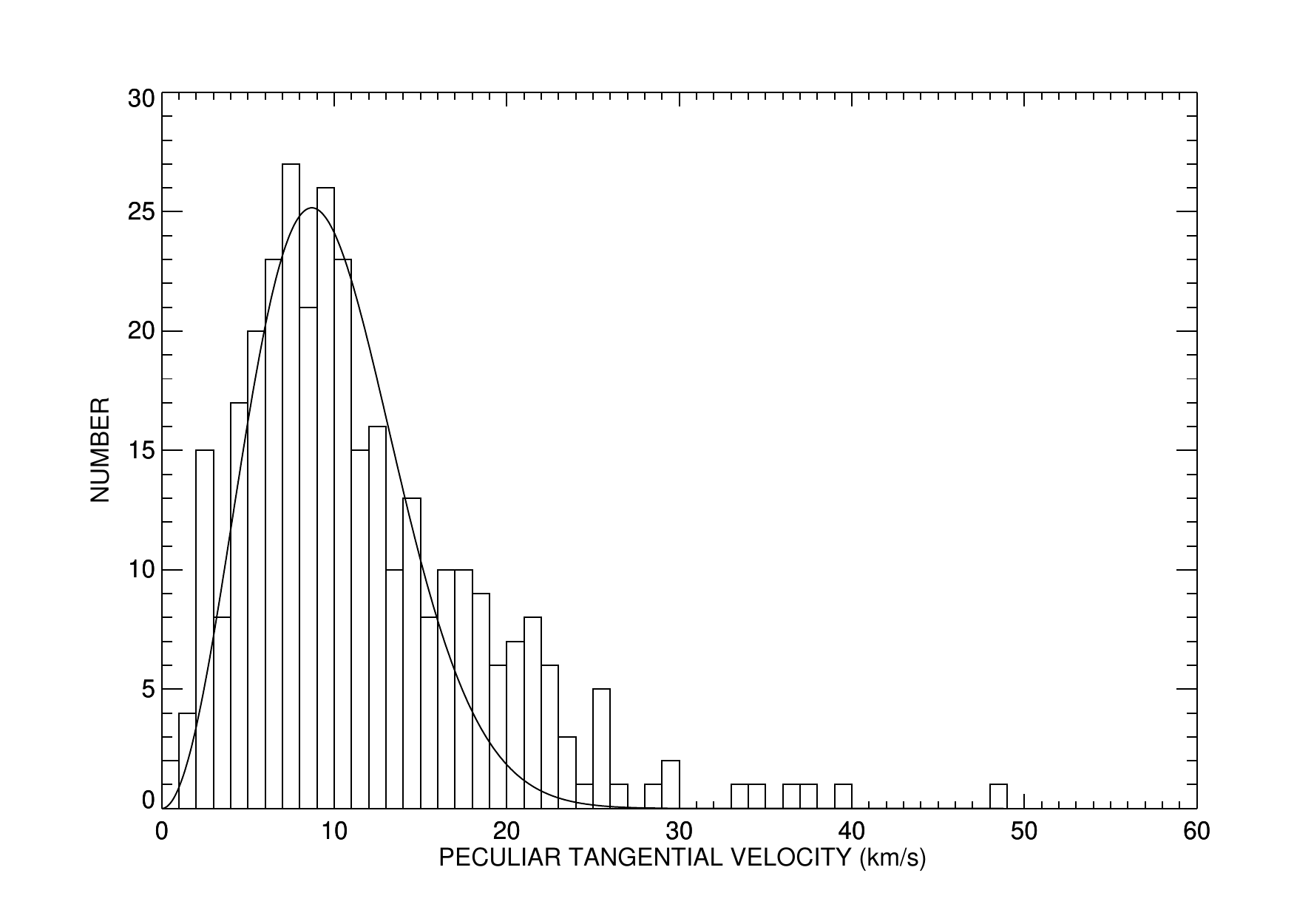}
\caption{The distribution of peculiar tangential velocities for 323 CBe stars identified from the LAMOST MRS DR7. A Maxwellian velocity distribution
was fitted to the peculiar tangential velocities. }
\label{fig:peculiar}
\end{figure*}

The OB runaway stars are stars with high spatial velocities that may travel far from the Galactic plane. \citet{Blaauw1961} suggested that the origin of 
OB runaway stars may be a result of binary supernovae scenario, in which the massive primary star of the binary system evolved to a supernova and
the lower mass secondary companion was released to become a high-velocity runaway star. 
Alternatively, \citet{Poveda1967} proposed that the runway stars may originate from a dynamic ejection mechanism as a consequence of 
gravitational interaction between stars in clusters. The origins and ejection mechanism for such runaways can be traced from studying the 
kinematics of the stars. Based upon the newly identified CBe stars from this work, by utilizing the astrometric solutions provided by the {\it Gaia} 
EDR3, we are interested in investigating the kinematic properties of the CBe stars by measuring the peculiar tangential velocities and searching for 
any possible runaway candidates. In order to measure the peculiar component velocities of the sample stars, we collected the available astrometric 
solutions for 323 CBe stars from the {\it Gaia} EDR3 and applied the quality filter following the same way as described in Section~\ref{sec:RVs}. We 
first transformed the equatorial coordinates of the sample stars to the Galactic coordinate systems by adopting the coordinate of the north Galactic 
pole of $(\alpha_{G}, \delta_{G}) = (192.95948^\circ, 27.12825^\circ)$ at the epoch of J2000.0 from the Hipparcos consortium. We then performed 
the matrix transformations for peculiar motions of the sample stars following the 
equations listed in \citet{Moffat1998,Moffat1999}. In order to subtract the contributions of the solar motion and the differential Galactic rotation from 
the peculiar motion of the CBe stars, we adopted the Solar motion of $(U_\odot, V_\odot, W_\odot) = (11.10, 12.24, 7.25)$ km s$^{-1}$ from 
\citet{Schonrich2010}. A flat rotation curve from \citet{Kerr1986} is adopted with the Solar galactocentric distance with value of $R_0 = 8.5$ kpc and 
the circular Galactic rotational velocity of $V_c = 220$ km s$^{-1}$. We plot the distribution of the peculiar tangential velocities for the CBe sample 
in Figure~\ref{fig:peculiar}. 

Stars with observed space velocities greater than 40 km s$^{-1}$ are classified as runaway stars \citep{Blaauw1961}. Many following works have 
adopted similar runaway criteria to search for runaway stars in the Galaxy either through measurements of the peculiar space motions or through 
the individual peculiar velocity components (radial or tangential components), such as using the Hipparcos astrometric observations to 
identify Galactic OB runaways \citep{Gies1986,Mdzinarishvili2005,Tetzlaff2011}, field O-type stars, and Wolf-Rayet stars 
\citep{Gies1987,Moffat1998,deWit2005}. \citet{Berger2001} utilized the {Hipparcos} measurements to identify runaway Be stars by adopting a 
selection criterion of peculiar space velocity greater than 40 km s$^{-1}$. Due to the absence of radial velocity measurements for the identified CBe 
stars in our sample, instead of adopting the historical runaway velocity limit, we thus rely on the peculiar tangential velocity to identify any potential 
runaway candidates. Because the distribution of peculiar tangential velocity includes contributions from both the longitudinal and latitudinal 
directions, a simple Gaussian is insufficient to describe such distribution. \citet{Hobbs2005} investigated the proper motion for a sample of pulsars and 
reported that a Maxwellian velocity distribution is well fitted to the space velocities of the stars. We then fitted a Maxwellian velocity distribution 
(Equation~\ref{eq:maxwellian}) to the peculiar tangential velocities,  
\begin{equation}
\label{eq:maxwellian}
f(v) = av^2 \exp(-v^2/b^2)
\end{equation}
where $a = 0.904 \pm 0.029$ and $b = 8.701 \pm 0.094$ km s$^{-1}$, and the distribution was overplotted in Figure~\ref{fig:peculiar}. We defined 
the criterion of runaway stars in our sample to stars with the peculiar tangential velocities greater than the value at $\sim1\%$ of the peak 
distribution from the Maxwellian fit, and this corresponds to a runaway velocity limit of $V_{Tp} > 24$ km s$^{-1}$. Based upon this proposed 
criterion for identifying the runaway stars, we identified 16 Be runaways from the sample of 323 stars. In 
Table~\ref{tab:peculiar}, we report the queried proper motions and parallax from the {\it Gaia} EDR3, measured peculiar tangential velocities 
($V_{Tp}$), and the associated uncertainties for the identified runaway candidate stars in columns 2 to 9. We also include the estimated heights of
the runaway stars from the Galactic plane in the last column of Table~\ref{tab:peculiar} adopting $z = r\sin{b}+z_\odot$, where $z_\odot$ with a value of 20.5 pc is taken from \citet{Humphreys1995}.          

We report a runaway rate of $\sim5\%$ for identified CBe stars from the LAMOST MRS DR7 sample, and this detection rate is comparable to the 
results reported by \citet{Berger2001}, who suggests a runaway frequency of $3\sim7 \%$ based upon an investigation of peculiar space motions 
for a sample of 344 Be stars. However, our result is significantly lower than the finding of $\sim13.1\%$ from \citet{Boubert2018}, in which a 
comprehensive collection of 632 Be stars were collected to investigate the runaway population of Be stars through a Bayesian approach. We
caution that the runaway fraction from the prior studies is based on measured peculiar space motion in both radial and tangential components. We 
may have missed the detections of runaways in our sample due to the absence of radial velocity measurements. We cross-matched the stars listed
in Table~\ref{tab:peculiar} to known runaways published in the literature, and no common stars are found, thus suggesting that these runaways are 
new identifications. 

\begin{figure*}[ht!]
\plotone{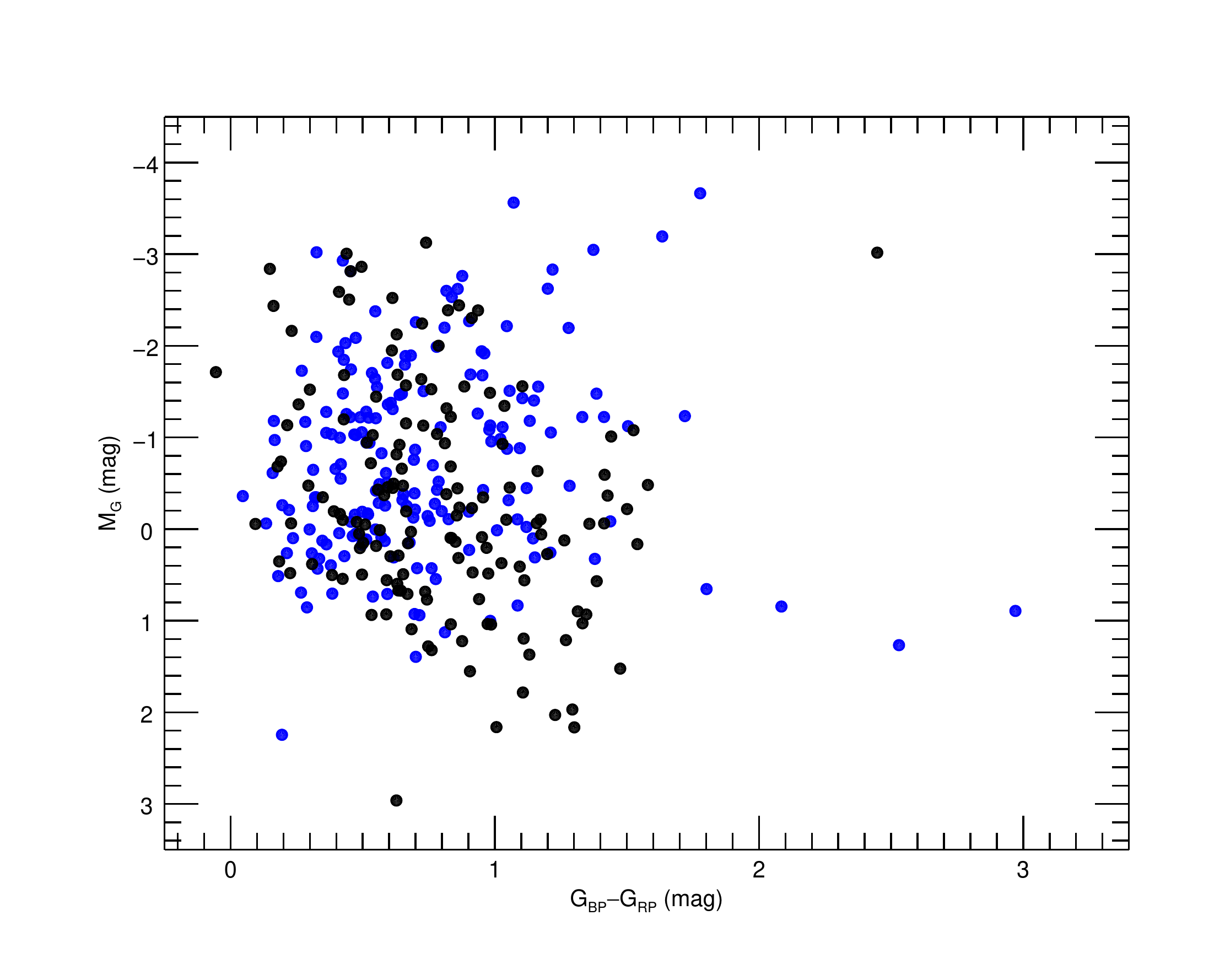}
\caption{The $G_{BP}-B_{RP}$ VS $G_{G}$ color-magnitude diagram for CBe stars identified from the LAMOST MRS DR7 database. The newly identified CBe stars from this work are plotted in blue and the prior known CBe stars cross-matched from literature are in black.   }
\label{fig:gaia_color}
\end{figure*}

Based upon the photometric observations for the CBe stars obtained from {\tt Gaia}, we show the location of both the newly identifications and literature cross-matched CBe stars on a $G_{BP}-G_{RP}$ VS $M_{G}$ color-magnitude diagram. We collected the photometric flux measurements from $G_{BP}$, $G_{RP}$, and $G$ bands from {\tt Gaia} EDR3. As suggested by \citet{gaia2018}, the flux excess factor (with a flag of {\tt phot\_bp\_rp\_excess\_factor}) within a range of $1.0+0.015\times(G_{BP}-G_{RP})^2 \le E \le 1.3+0.06\times(G_{BP}-G_{RP})^2$ was checked for each observation to reject any excess flux recorded in the $G_{BP}$ and $G_{RP}$ bands. We further adopted the approach from \citet{Luri2018} to convert the observed apparent $G$ magnitude to absolute $M_G$ magnitude (see their Equation 22). In Figure~\ref{fig:gaia_color}, we show the $G_{BP}-G_{RP}$ VS ${M_G}$ color-magnitude plot for 179 new identifications (blue) and literature cross-matched 149 CBe stars (black) with available measurements from {\tt Gaia}. The CBe stars are located on the color-magnitude diagram as expected for Be stars.

\begin{deluxetable*}{llcccccccccc}
\rotate
\tabletypesize{\tiny}
\tablecaption{Open cluster memberships for identified CBe stars from the LAMOST MRS DR7 sample \label{tab:OC}}
\tablewidth{0pt}
\tablehead{
\colhead{LAMOST}  & \colhead{Cluster}   & \colhead{Cluster $V_r$}  & \colhead{$\sigma (V_r)$} & \colhead{Distance} &\colhead{$\sigma (d)$}  & \colhead{$\log{Age}$}  & \colhead{$\sigma (Age)$} & \colhead{$[Fe/H]$} & \colhead{$\sigma ([Fe/H])$} & \colhead{$A_V$}  & \colhead{$\sigma (A_V)$} \\
\colhead{designation}  &  \colhead{Name} & \colhead{(km s$^{-1}$)} & \colhead{(km s$^{-1}$)}  & \colhead{(pc)} & \colhead{(pc)} & \colhead{(yr)} &  \colhead{(yr)} &\colhead{} & \colhead{} &\colhead{(mag)} &\colhead{(mag)}
}
\startdata
J011524.01+583108.5	& 	 NGC 457  &  \phn$-$4.6 & 1.1	& 	2540	& 	133	& 	7.373 	& 0.073	& 	$-$0.034 	& 	0.153	& 	1.612	& 	0.020 \\ 
J011540.89+584902.1	& 	  NGC 436 	&  $-$79.8 	& 0.4 	& 2743 	& 118 	& 7.861 	& 0.105 	& $-$0.122 	& 0.110 	& 1.590  	&0.034 \\ 
J011604.49+584651.2	& 	  NGC 436	& $-$79.8 	& 0.4 	& 2743 	& 118  	&7.861 	& 0.105 	& $-$0.122 	& 0.110  	&1.590 	& 0.034  \\
J011833.06+582230.4	& 	  NGC 457 	& \phn$-$4.6 	& 1.1 	& 2540  	&133 	& 7.373 	& 0.073 	& $-$0.034 	& 0.153 	& 1.612 	& 0.020 \\
J011854.11+581206.8	& 	  NGC 457 	& \phn$-$4.6  	&1.1  	&2540  	&133  	&7.373  	&0.073  	&$-$0.034  	&0.153  	&1.612  	&0.020 \\
J011902.35+581920.2 	& 	 NGC 457 	& \phn$-$4.6  	&1.1  	&2540  	&133  	&7.373  	&0.073  	&$-$0.034  	&0.153  	&1.612  	&0.020 \\ 
J014430.77+584610.9	& 	  COIN-Gaia 4	& \nodata & \nodata 	&1853  	&146  	&8.238  	&0.318  	&$-$0.104  	&0.103  	&1.767  	&0.134 \\
J015822.85+552725.9	& 	  NGC 744	& \nodata &\nodata 	&1258  	& \phn40  	&8.261  	&0.249  	& \phs0.019  	&0.076  	&1.116  	&0.107 \\
J021346.59+600208.3	& 	  COIN-Gaia 35 & \nodata & \nodata &2354 	& 290  	&7.013  	&0.616  	&$-$0.140  	&0.174  	&2.726  	&0.091 \\
J021547.53+572514.2	& 	 UBC 46 		&\nodata	&\nodata  		&2003  	&\phn77 		&7.613 	&0.140 	&$-$0.143 	&0.072 	&1.762	& 0.033 \\
J022142.93+570530.6	& 	 NGC 884 	&  \nodata &\nodata           	&2150 	&103 	&7.187 	&0.050 	&$-$0.079 	&0.092 	&1.709 	&0.019 \\
J022143.39+570732.8	& 	 NGC 884  	&  \nodata &\nodata            	&2150 	&103 	&7.187 	&0.050 	&$-$0.079 	&0.092 	&1.709 	&0.019 \\
J022144.47+571052.2	& 	 NGC 884         & \nodata  & \nodata    		& 2150 	&103 	&7.187 	&0.050 	&$-$0.079 	&0.092 	&1.709 	&0.019 \\
J022145.96+570500.9	& 	 NGC 884		& \nodata  & \nodata    		& 2150 	&103 	&7.187 	&0.050 	&$-$0.079 	&0.092 	&1.709 	&0.019 \\
J022202.47+570920.4	& 	 NGC 884		& \nodata  & \nodata    		& 2150 	&103 	&7.187 	&0.050 	&$-$0.079 	&0.092 	&1.709 	&0.019 \\
J022204.56+571038.8	& 	 NGC 884		& \nodata & \nodata    		& 2150 	&103 	&7.187 	&0.050 	&$-$0.079 	&0.092 	&1.709 	&0.019 \\
J022246.97+565805.8	& 	 NGC 884 	&   \nodata  & \nodata    		& 2150 	&103 	&7.187 	&0.050 	&$-$0.079 	&0.092 	&1.709 	&0.019 \\
J022248.96+570913.9	& 	 NGC 884 	& \nodata   &  \nodata   		& 2150 	&103 	&7.187 	&0.050 	&$-$0.079 	&0.092 	&1.709 	&0.019 \\
J022250.28+570850.6	& 	 NGC 884 	& \nodata  &\nodata     		& 2150 	&103 	&7.187 	&0.050 	&$-$0.079 	&0.092 	&1.709 	&0.019 \\ 
J022304.18+570738.7	& 	 NGC 884 	& \nodata  & \nodata    		& 2150 	&103 	&7.187 	&0.050 	&$-$0.079 	&0.092 	&1.709 	&0.019 \\
J022324.92+571903.3	& 	 NGC 884 	& \nodata   & \nodata    		& 2150 	&103 	&7.187 	&0.050 	&$-$0.079 	&0.092 	&1.709 	&0.019 \\ 
J022328.16+572325.6 	& 	NGC 884  	& \nodata   & \nodata    		& 2150 	&103 	&7.187 	&0.050 	&$-$0.079 	&0.092 	&1.709 	&0.019 \\
J022432.59+570044.8	& 	 NGC 884  	& \nodata  & \nodata    		& 2150 	&103 	&7.187 	&0.050 	&$-$0.079 	&0.092 	&1.709 	&0.019 \\
J022650.54+572020.1 	& 	UBC 192   	& \nodata    &\nodata              & 2239  	&\phn75 	&7.278 	&0.176  	& \phs0.021 	&0.072 	&1.773 	&0.036 \\ 
J023005.94+571833.8 	& 	UBC 606		 & \nodata		&\nodata    &2463 	& \phn54	& 7.000	& 0.236	&$-$0.127		&0.133	&2.168	&0.036 \\ 
J023058.70+571503.8 	& 	UBC 192 		&\nodata		&\nodata              &2239  	&\phn75	& 7.278 	&0.176	& \phs0.021	&0.072	&1.773	&0.036 \\ 
J050317.28+234917.4 	& 	NGC 1750 	& \phn$-$9.7 &1.1  	&\phn707 	&  \phn\phn7 		&8.296	&0.196	& \phs0.060		&0.057	&1.239	&0.034 \\ 
J053607.20+340803.2 	& 	NGC 1960 	&\nodata 		&\nodata    		&1086  	&\phn37		&7.480	&0.059 	&$-$0.030		&0.085	&0.929 	&0.024 \\ 
J054946.16+285952.8	& 	 Czernik 23 	& \phs14.3	& 0.5 	&3102 	&221 	&8.495	&0.617	&$-$0.132		& 0.167 	&1.743	&0.155 \\ 
J060452.44+240331.3 	& 	IC 2157  		&\nodata 		&\nodata 		&1885 	&155 	&7.552	&0.185 	&$-$0.098 	&0.055 	&1.651 	&0.051 \\ 
J061410.98+123607.0 	& 	NGC 2194 	& \phs41.6	& 1.4 	&2869 	&\phn83 		&8.823 	&0.050 	&$-$0.067 	&0.066 	&1.562 	&0.075 \\
J062140.49+265806.2 	& 	Gulliver 56   	& \nodata              &	\nodata	&1943  	&\phn51 		&8.350 	&0.174 	&$-$0.100 	&0.047 	&0.968 	&0.076 \\ 
J063131.81+053051.6	& 	 NGC 2244   	& \phs92.3 	&1.2	& 1254  	&\phn89 		&7.111 	&0.090 	&$-$0.214 	&0.084 	&1.571 	&0.063 \\ 
J063337.49+044847.0 	& 	NGC 2244	& \phs92.3	& 1.2 &1254 & \phn89 &7.111 &0.090 &$-$0.214 & 0.084 &1.571 &0.063 \\ 
J063701.44+051307.0 	& 	Collinder 107 	&\nodata 		& \nodata             &1466  	&\phn85 		&7.183 	&0.129 	&$-$0.100 	&0.163 	&1.465 	&0.071 \\ 
J184244.50$-$060913.6 	& 	UBC 104   	& \nodata		&\nodata             	&2741 	&151 	&8.428 	&0.634 	& \phs0.295	&0.112 	&1.781 	&0.154 \\
J185103.91$-$061732.0 	& 	NGC 6705 	& \phs35.7 	& 0.2 	&1888 	&\phn65 		&8.469 	&0.085  	& \phs0.043 	&0.063 	&1.457 	&0.051 \\ 
J221255.97+572639.0	& 	 NGC 7235  	&\nodata 		&\nodata      		&3158	& 261 	&7.100 	&0.117  	& \phs0.018	& 0.115 	&2.566 	&0.031 \\
J221950.08+572428.5	& 	LP 1809  		& $-$84.3 &3.1 &2707 	&191 	&8.644 	&0.645  	& \phs0.092 	&0.116 	&2.729 	&0.191 \\ 
J221951.45+580853.4	& 	 NGC 7261  	& \nodata		&\nodata		&2389 	&106 	&7.632 	&0.088 	& \phs0.010 	&0.096 	&2.731 	&0.029 \\ 
J235039.51+605439.1	& 	 FSR 0451	&\nodata  		&\nodata		&2602 	&120 	&7.077 	&0.024 	&$-$0.095 	&0.063 	&2.291 	&0.045 \\ 
 \enddata
\tablecomments{This table is available in its entirety in machine-readable form.}
\end{deluxetable*}

\begin{deluxetable*}{lccccccccc}
\tablecaption{The peculiar velocities of the 16 runaway CBe stars identified from the LAMOST MRS DR7 sample \label{tab:peculiar}}
\tablewidth{0pt}
\tablehead{
\colhead{LAMOST}  & \colhead{$\mu_\alpha$}   & \colhead{$\sigma$}  & \colhead{$\mu_\beta$} & \colhead{$\sigma$} &\colhead{Parallax}  & \colhead{$\sigma$}  & \colhead{$V_{Tp}$} & \colhead{$\sigma$} & \colhead{$z$} \\
\colhead{designation} & \colhead{(mas yr$^{-1}$)} & \colhead{(mas yr$^{-1}$)} & \colhead{(mas yr$^{-1}$)} & \colhead{(mas yr$^{-1}$)} & \colhead{(mas)} & \colhead{(mas)} & \colhead{(km s$^{-1}$)} & \colhead{(km s$^{-1}$)} & \colhead{(pc)}
}
\startdata
J032822.23+450756.0	& \phs2.4793	&  0.0240& $-$0.0143& 0.0194& 0.3898&  0.0209& 34.3&   1.9&     212.6\\
J041706.17+490757.2&    \phs0.7924&    0.0172&   $-$2.3405&   0.0145&    0.3515&    0.0172&      24.7&     1.2&      \phn31.5\\
J044532.80+441554.7	& \phs4.1211	&  0.0231& $-$6.8914& 0.0172& 0.7670&  0.0174& 36.7&   0.9&      \phn44.8\\
J044800.93+401847.9&    \phs0.1580&    0.0193&   $-$0.8861&   0.0149&    0.1140&    0.0150&      26.2&       3.6&      \phn82.6\\
J055812.33+274333.7&   $-$0.8517&    0.0247&    \phs0.5285&   0.0173&    0.3135&    0.0208&      25.0&       1.7&      \phn94.8\\
J060421.79+283840.7&    \phs0.8418&    0.0245&   $-$2.7897&   0.0174&    0.3663&    0.0223&      25.1&       1.6&      \phn62.1\\
J061358.64+124327.3	& $-$0.2615	& 0.0162& $-$3.2853& 0.0130& 0.4242&  0.0157& 29.8&  1.1&      \phn41.6\\
J062945.80+173216.5	& \phs0.8607	&  0.0339& $-$2.4530& 0.0259& 0.3150&  0.0292& 33.4&   3.2&     170.0\\
J185103.91$-$061732.0&   $-$1.5166&    0.0236&   $-$4.1254&   0.0193&    0.3748&    0.0264&      25.9&   1.9&     100.5\\
J221950.08+572428.5	& $-$3.6236	& 0.0149& $-$2.8681& 0.0142& 0.2148&  0.0134& 29.1&  1.9&     154.4\\
J221951.45+580853.4&   $-$3.8237&    0.0119&   $-$3.1010&   0.0123&    0.2858&    0.0115&      25.4&    1.1&     192.2\\
J222236.47+531945.9	& $-$4.5890	& 0.0188& $-$3.7029& 0.0173& 0.2318&  0.0149& 48.8&  3.2&     123.3\\
J222426.67+573834.4	& $-$3.8290	& 0.0134& $-$3.7036& 0.0116& 0.3252&  0.0119& 28.4&  1.1&     177.2\\
J223529.46+564759.7	& $-$3.7224	& 0.0148& $-$3.0935& 0.0138& 0.2040&  0.0139& 39.4&  2.7&     161.1\\
J234525.42+564015.2&   $-$1.5749&    0.0150&    \phs0.9009&   0.0153&    0.4647&    0.0160&      25.3&    0.9&      \phn62.4\\
J235756.00+625549.6	& $-$4.2501	& 0.0137& $-$1.1622& 0.0144& 0.2653&  0.0136& 37.7&  2.0&     295.8\\
\enddata
\tablecomments{This table is available in its entirety in machine-readable form.}
\end{deluxetable*}

\section{Summary and Conclusions} \label{sec:conc}
The classical Be stars are B-type main-sequence stars surrounded by circumstellar disks that create Balmer emission lines in their spectra. Here we examined the large and homogenous spectroscopic observations in the LAMOST MRS DR7 survey, we use a deep convolutional neural 
network, {\tt ResNet} to identify CBe stars from their H$\alpha$ properties. 

Our first task was to form an initial sample of early-type stars (with the spectral classification of O-type to early A-type only) from a collection of 
2,260,387 spectra for 789,918 stars from the database. We rejected 1,973,886 spectra for late A-type and F-type stars from the sample using the 
estimated $T_{\rm eff}$ values as given from the \emph{LASP}. Through calculations of the equivalent widths of H$\alpha\ \lambda6563$ and
\ion{Mg}{1\ $b$} triplet line profiles, by forming a distribution of $EW$ over the two selected line indices, we excluded any hidden late-type stars and 
narrowed the sample down to 261,647 spectra. Further filtering of K-type and M-type stars was achieved by inspecting their locations on a color-color
diagram using the IR photometric observations from the \emph{2MASS} all-sky survey. We selected a collection of 257,860 early-type
stellar spectra from the MRS DR7 database.

In order to apply the {\tt ResNet} network to identify Be candidate stars from the initial sample of 257,860 early-type stellar spectra, we first adopted a Bartlett window smoothing approach to smooth observed spectra over a wavelength region of $6551-6579$ \AA\ to identify the H$\alpha$ peak features in the sample. Through visual inspection, we constructed a training sample of 2,084 spectra, in which 1,042 of them were found as Be candidate spectra showing H$\alpha$ peak feature in their profiles, and the rest of the spectra were randomly 
chosen from the initial sample as non-Be spectra. By applying the {\tt ResNet-18} learning module and then confirming the identification with a visual inspection, we arrived at a preliminary list of 1,162 Be stars identified from the DR7 database. The learning module achieved a classification accuracy of 99.5\%.

We cross-matched these candidate stars with published catalogs and found that 151 of them are known CBe stars. By further applying a three-step 
test, 183 new detections were identified from the sample. Based upon cross-matching the equatorial coordinates, proper motion, and parallax of the 
identified CBe stars with those of known open clusters, we identified 41 cluster member stars. The parent clusters of these member stars have ages 
between 10$\sim$655 Myr, with metallicity $[Fe/H]$ spans a range of $-$0.214$\sim$0.295. We also explored the kinematics of the identified CBe stars by measuring their peculiar tangential velocities based upon astrometric solutions from the \emph{Gaia} EDR3 and identified 16 new 
runaways from the sample. These newly identified runaways correspond to a $\sim$5$\%$ detection rate, which is comparable to the findings reported from earlier studies made from the \emph{Hipparcos} observations. These runaways have measured Galactic heights between 30$\sim$300 pc from the Galactic disc, no high-latitude CBe stars were found in the sample. Both the dynamic ejection mechanism and the binary supernova scenario suggest that the formation of runaways is associated with binary systems. \citet{Leonard1990} performed an N-body simulation for the former scenario and estimated a binary fraction of $10\%$ for runaways. In contrast, a binary frequency of 30$\sim$40$\%$ was reported from \citet{Portegies2000} by utilizing binary population synthesis calculations for the binary supernova scenario. Future investigation for the binary properties of these newly identified runaways from the LAMOST MRS sample will be helpful to constrain their formation channel. 

In this work, we have demonstrated the applicability of applying the neural network {\tt ResNet} to identify the spectral features of CBe stars from a large collection of observations efficiently and accurately. We have built the Be spectra classifier using the {\tt ResNet} module for the observations from the LAMOST MRS DR7 through the identification of H$\alpha$ peak featured spectra and the construction of the training sample. In our future work, we plan to include the spectral processing process into the classifying scheme, such as the spectral rectification, and release the code as a public accessible package. By providing input spectra obtained from the database, such a classifier will automatically rectify the spectra and perform the classification task to further improve the identification efficiency. Because of the limited wavelength coverage of the MRS spectra, we solely relied
on the spectra recorded in the red arm covering the H$\alpha$ profile to perform the identification task. We are optimized about the future work of applying this technique to the low-resolution spectra (LRS) from the LAMOST database since such spectra display a broader optical wavelength coverage to include the major atmospheric line profiles of Be stars compared to the MRS spectra. The extensive collection of spectroscopic observations made from the LAMOST MRS DR7 database comprises the first 
of the five-year time-domain observing campaign, with the future spectra from the forthcoming DR8 and DR9, we would expect to enlarge the 
current sample size of CBe stars. This information will provide a reference for future investigations of the Be population and its physical and evolutionary 
properties.

\begin{acknowledgments}
This work was supported by the Chinese National Science Foundation under programs of No.\ 12090040, 12090043, 11733008, and 12103064.\ 
Guoshoujing Telescope (the Large Sky Area Multi-Object Fiber Spectroscopic Telescope LAMOST) is a National Major Scientific Project built by the 
Chinese Academy of Sciences.\ Funding for the project has been provided by the National Development and Reform Commission.\ 
LAMOST is operated and managed by the National Astronomical Observatories, Chinese Academy of Sciences.\ This work has made use of data 
from the European Space Agency (ESA) mission {\it Gaia} (\url{https://www.cosmos.esa.int/gaia}), processed by the {\it Gaia} Data Processing and 
Analysis Consortium (DPAC, \url{https://www.cosmos.esa.int/web/gaia/dpac/consortium}).\ Funding for the DPAC has been provided by national 
institutions, in particular the institutions participating in the {\it Gaia} Multilateral Agreement.\ This work has made use of the BeSS database,
operated at LESIA, Observatoire de Meudon, France: http://basebe.obspm.fr.   
\end{acknowledgments}

%

\vspace{5mm}
\facilities{LAMOST, Gaia}





\appendix

\section{Morphological classification of the H$\alpha$ shell line spectra}
\begin{figure*}[ht!]
\plotone{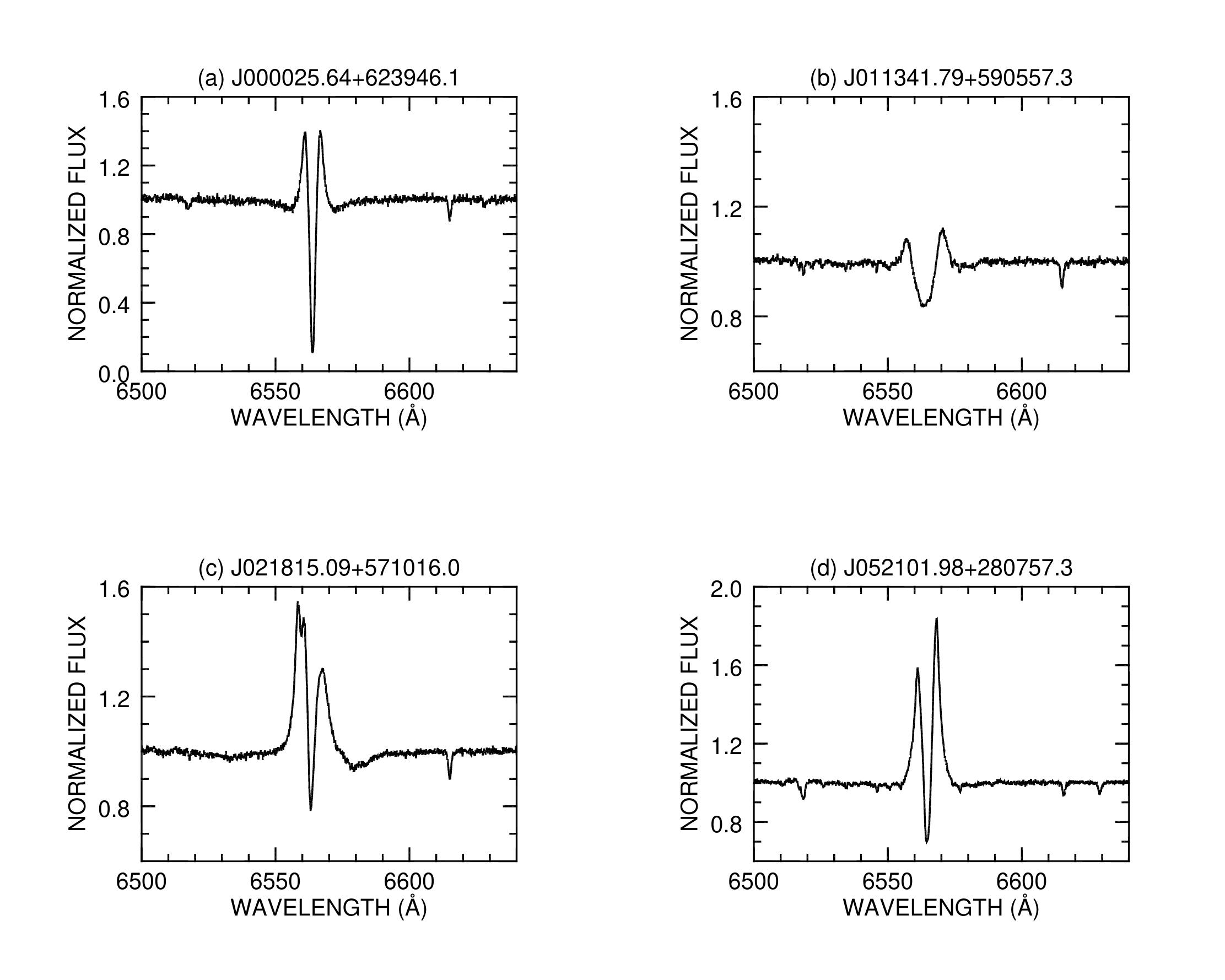}      
\caption{The H$\alpha$ emission profiles superimposed with an absorption shell line identified from the LAMOST MRS DR7 database. (a) J000025.64+623946.1 (Type 1), (b) J011341.79+590557.3 (Type 2), (c) J021815.09+571016.0 (Type 3), and (d) J052101.98+280757.3 (Type 3).}
\label{fig:emission-shell}
\end{figure*}

Shell line features appear in the H$\alpha$ profile of 312 Be candidate stars. The red spectra of 113 Be candidate stars display a superimposed absorption shell line on
top of the H$\alpha$ emission profile. We classified these emission shell line profiles into three sub-categories: a spectrum with symmetric narrow absorption shell profile 
denoted as Type 1, a spectrum with symmetric broad absorption shell profile (Type 2), and an asymmetric absorption shell (Type 3). In Figure~\ref{fig:emission-shell}, 
we show the example spectra of Type 1 and Type 2 in panels (a) to (b), respectively. An asymmetric absorption shell line superimposed on top of the H$\alpha$ emission 
profile of Type 3 spectra is shown in panels (c) and (d). We report the LAMOST designation ID, right ascension (RA), declination (DEC) in the first three columns of 
Table~\ref{tab:emission-shell}, and list the shell line type for each of the stars in the last column of the table.

\begin{figure*}[ht!]
\plotone{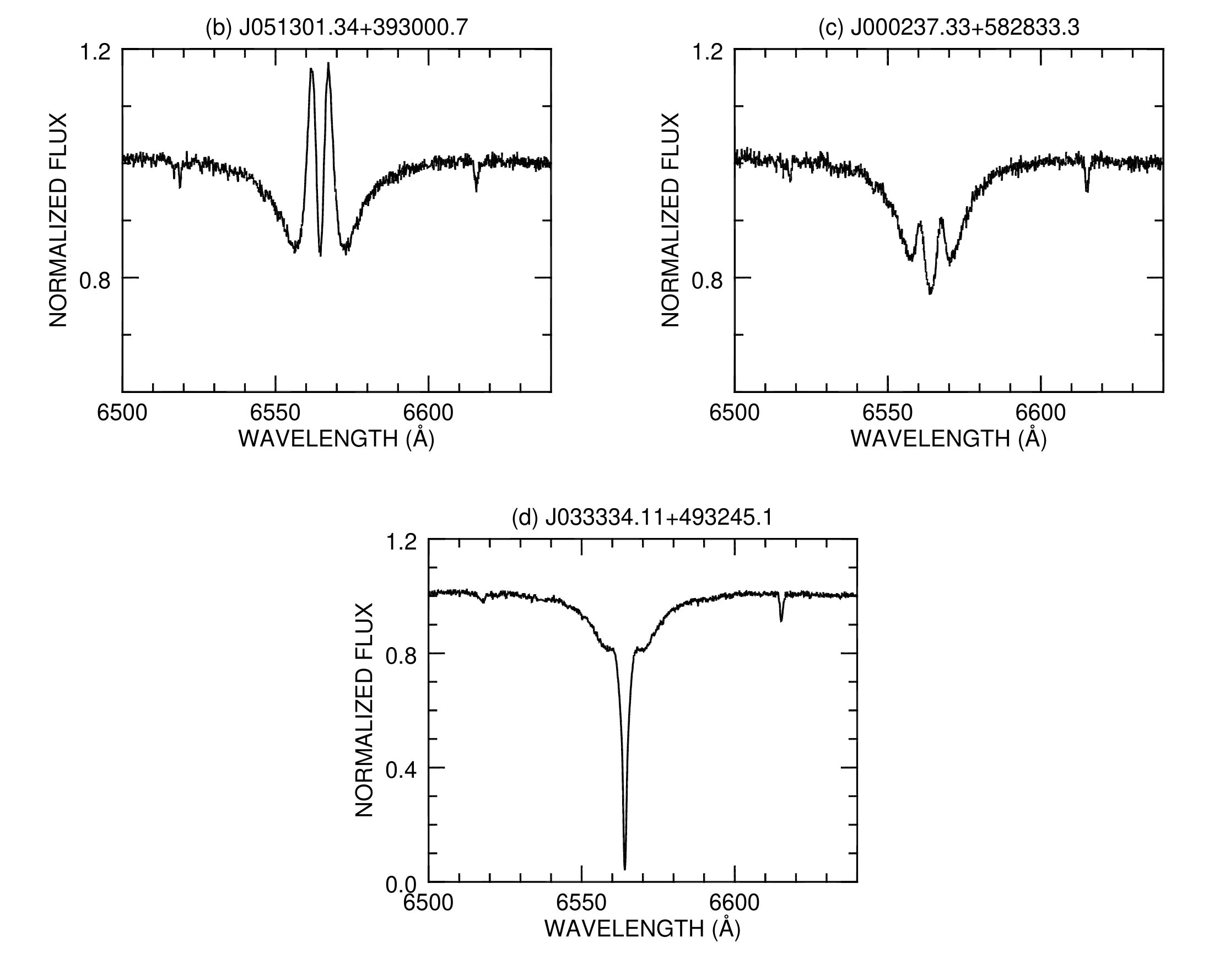}   
\caption{The H$\alpha$ absorption profiles superimposed with a shell line identified from the LAMOST MRS DR7 database. (a) J051301.34+393000.7 (Type 1), (b) J000237.33+582833.3 (Type 2), and (c) J033334.11+493245.1 (Type 3).}
\label{fig:absorption-shell}
\end{figure*}

Shell line features were superimposed on top of a broad H$\alpha$ absorption profile for the rest of the 199 shell stars. In Figure~\ref{fig:absorption-shell}, we show examples of
shell line spectra with an emission component above the spectral continuum (Type 1, panel a), an emission component below the spectral continuum (Type 2, panel b), and a 
narrow absorption shell line (Type 3, panel c). We classified these stars into sub-types based upon the shell line classification as mentioned through visual inspection. The 
equatorial coordinates and the classification types are collected in Table~\ref{tab:absorption-shell}, in the same format as 
Table~\ref{tab:emission-shell}.    

\placefigure{fig:Emission-shell}
\restartappendixnumbering

\begin{deluxetable*}{lccc}\label{tab:emission-shell}
\tabletypesize{\scriptsize}
\tablecaption{Candidate Be stars displaying H$\alpha$ emission profiles with absorption shell line identified from the LAMOST MRS DR7 sample \label{tab:OCs}}
\tablewidth{0pt}
\tablehead{
\colhead{LAMOST}  & \colhead{RA}   & \colhead{DEC} & \colhead{Type Index} \\
\colhead{designation}  &  \colhead{(deg)} & \colhead{(deg)} & \colhead{}
}
\startdata
J000025.64+623946.1 &   0.1068 & 62.6628 &1 \\
J000336.20+554905.9 &  0.9008  &55.8183 &3  \\
J000719.05+615350.5 &  1.8294 & 61.8974 &1  \\
J004746.45+601122.5 & 11.9436 &60.1896 &1  \\
J011341.79+590557.3 & 18.4241 &59.0993 &2  \\
J011507.07+581812.5 &  18.7795 &58.3035 &3  \\
J011623.65+590507.3 & 19.0986 &59.0854 &1  \\
J011623.65+590507.4 & 19.0986 &59.0854 &1  \\
J012318.25+580435.1 &  20.8261& 58.0764 &3  \\
J013131.58+594607.5 & 22.8816 &59.7688 &2  \\
\enddata
\tablecomments{This table is available in its entirety in machine-readable form. The first ten entries are shown here for guidance regarding
its format and content.}
\end{deluxetable*}

\placefigure{fig:Absorption-shell}

\begin{deluxetable*}{lccc}\label{tab:absorption-shell}
\tabletypesize{\scriptsize}
\tablecaption{Candidate Be stars displaying H$\alpha$ absorption profiles with shell line identified from the LAMOST MRS DR7 sample \label{tab:OCs}}
\tablewidth{0pt}
\tablehead{
\colhead{LAMOST}  & \colhead{RA}   & \colhead{DEC} & \colhead{Type Index} \\
\colhead{designation}  &  \colhead{(deg)} & \colhead{(deg)} & \colhead{}
}
\startdata
J000021.08+552725.1 &  0.0878 &55.4570 &2  \\
J000056.27+625702.5  & 0.2345 &62.9507 &2  \\
J000237.33+582833.3  & 0.6556 &58.4759 &2  \\
J000452.79+623619.8  & 1.2200 &62.6055 &1  \\
J000805.45+612513.9  & 2.0227 &61.4205 &3  \\
J002922.13+580036.7  & 7.3422 &58.0102 &3  \\
J003150.41+573524.4  & 7.9601 &57.5901& 2  \\
J003504.28+604317.5  & 8.7678 &60.7216 &2  \\
J011242.62+595723.1 & 18.1776 &59.9564 &1  \\
J011249.59+593434.2 & 18.2066 &59.5762 &1 \\
\enddata
\tablecomments{This table is available in its entirety in machine-readable form. The first ten entries are shown here for guidance regarding
its format and content.}
\end{deluxetable*}

\bibliography{ms.bib}{}
\bibliographystyle{aasjournal}



\end{document}